\documentclass[reprint,amsmath,amssymb,aps]{revtex4-1}
\usepackage[colorlinks,citecolor=blue,linkcolor=blue,anchorcolor=blue,filecolor=blue,
urlcolor=blue]{hyperref}
\usepackage{graphicx}
\usepackage{ulem}
\usepackage{dcolumn}
\usepackage{bm}
\usepackage{todonotes}

\begin{document}

\title{PNJL model at zero temperature: three-flavor case}

\author{O. A. Mattos$^1$, T. Frederico$^1$, C. H. Lenzi$^1$, M. Dutra$^1$, and O. Louren\c{c}o$^1$}  
\affiliation{Departamento de F\'isica, Instituto Tecnol\'ogico de Aeron\'autica, DCTA, 12228-900, S\~ao Jos\'e dos Campos, SP, Brazil}

\date{\today}

\begin{abstract}
We propose a three-flavor version of the Polyakov-Nambu-Jona-Lasino (PNJL) model at zero temperature regime, by implementing a traced Polyakov loop ($\Phi$) dependence in the scalar, vector and 't Hooft channel strengths. We study the thermodynamics of this model, named as PNJL0, with special attention for the first order confinement/deconfinement phase transition for which $\Phi$ is the order parameter. For the symmetric quark matter case, an interesting feature observed is a strong reduction of the constituent strange quark mass ($M_s$) at the chemical potential related to point where deconfinement takes place. The emergence of $\Phi$ favors the restoration of chiral symmetry even for the  strange quark. We also investigate the charge neutral system of quarks and leptons in weak equilibrium. As an application, we construct a hadron-quark phase transition with a density dependent hadronic model coupled to the SU(3) PNJL0 model. In this case, the quark side is composed by deconfined particles. This approach is used to determine mass-radius profiles compatible with recent data from the Neutron Star Interior Composition Explorer (NICER) mission.
\end{abstract}


\maketitle

\section{Introduction} 

The  understanding of the physics related to particles interacting strongly is a fundamental issue studied for both, theoretical and experimental researchers. The investigations of collisions at ultrarelativistic energies in the most sophisticated accelerators of the world, such as the Relativistic Heavy-Ion Collider (RHIC)~\cite{rhic} and the Large Hadron Collider (LHC)~\cite{lhc}, for instance, give  the support needed for the study of the strong force probed at the experimental level. The forthcoming Electron Ion Collider (EIC)~\cite{eic} at Brookhaven, the
Facility for Antiproton and Ion Research (FAIR)~\cite{fair}, at Darmstadt, and the Nuclotron-based Ion Collider fAcility (NICA)~\cite{nica}, at the Joint Institute for Nuclear Research (JINR), in Dubna, are examples of other facilities with the same purpose. Many aspects of the strongly interacting matter can be directly or indirectly accessed from experiments, as  for example, the charge-parity violation (CPV) observed in heavy mesons decays. The reader is addressed to Ref.~\cite{bediaga} and references therein for an overview including the role of the strong force in forming the observed CPV pattern.

From the theoretical point of view, on the other hand, the physics of quarks and gluons is described by Quantum Chromodynamics (QCD)~\cite{qcd1,qcd2,qcd3}. Its nonperturbative regime, i.e., QCD at large distances, or equivalently, low energies, can be investigated through the lattice discretization of QCD in the Euclidean space~\cite{kogut1,kogut2,rothe} where the numerical calculations are based on Monte Carlo simulations~\cite{landau}. However, the so called fermion sign problem~\cite{sign} restricts such technique when quarks at finite chemical potential are included in the system. A second approach applied to study the infrared region of QCD is the use of continuous methods relying on Dyson-Schwinger equations (DSE) in Euclidean space~\cite{dse}, used to determine the equations of motion of the $n$-point functions. Another practical method of treating quarks and gluons, in systems in which such particles are the fundamental degrees of freedom, is the use of phenomenological models. Probably, the simplest one used for this aim is the famous MIT bag model~\cite{mit}. In its original version, it establishes that quarks are inside a bag that effectively corresponds to a confining potential. The constant $B$ is added to the free quark pressure, and all thermodynamics is based on this simple assumption. 

Other widely used effective quark model is the Nambu-Jona-Lasinio (NJL) one~\cite{Nambu1,Nambu2,buballa,vogl2,outros}. It was proposed by Y. Nambu and G. Jona-Lasinio before QCD and originally described nucleons and pions. In the context of the quark interaction, it presents some similarities with QCD such as the dynamical breaking of chiral symmetry, and the associated quark mass generation due to the quark condensates. Similar nonperturbative phenomena also occur in the Bardeen-Cooper-Schrieffer (BCS) model of superconductivity~\cite{bcs} (in which the NJL model was inspired), where interactions give rise to energy gaps. In the NJL model, the ``gap'' in the quark mass is produced by the point-like interactions between Dirac particles. 

Although capable of describing some important features of QCD, the NJL model fails to reproduce a particular phenomenon, namely, the infrared dynamics of confinement as well as the deconfinement of quarks, related to the so called ``asymptotic freedom'', discovered by Gross, Wilczek and Politzer~\cite{gross}. In the NJL model there is no information regarding the interaction among the gluons. Such an issue was corrected by K. Fukushima~\cite{FUKUSHIMA}, who proposed the inclusion of these mediators through a background field obtained from the trace of a matrix in color space associated to the gluon gauge field $A_4$. The so-called traced Polyakov loop, $\Phi$ (and its conjugate $\Phi^*$), mimics the gluon dynamics in an effective way. Physically, $\Phi$ can be seen as the order parameter of the confinement/deconfinement phase transition, since it is related to $e^{-F/T}$, with $F$ being the free energy of an external static quark. In the confined phase, the free energy of a single quark is infinite and, consequently, $\Phi=0$. In the deconfined phase, on the other hand, $F$ is finite and $\Phi\ne 0$. However, we remind that $\Phi$ is not enough to have a true confinement in the PNJL model at finite temperature since it only presents a statistical confinement 
while asymptotic quark states still exists.
The aforementioned analysis originally does not apply at zero temperature in the PNJL model and in most of its parametrizations, since in this case the Polyakov potential vanishes and the modified Fermi-Dirac distributions reduces to the step functions. Because of that, confinement effects are lost  and it is not possible to investigate high density deconfined quark systems at $T=0$. In order to circumvent this issue, it was proposed in Refs.~\cite{mattos,epjc} a modification of the strengths of the couplings in the NJL model by making them dependent on $\Phi$. As a direct consequence, the Polyakov loop potential became nonvanishing even at $T=0$. In this way, the new model constructed for flavor SU(2) quarks, named as PNJL0 model, was able to exhibit the confinement/deconfinement phenomenology at $T=0$. In this work, we extend these previous studies by generalizing the PNJL0 model to include flavor SU(3) quarks. We investigate its thermodynamics at $T=0$ and also perform an application in the context of the hadron-quark phase transition.

The paper is organized as follows. In Sec.~\ref{secpnjl} we review the main equations of the traditional PNJL at finite temperature. In Sec.~\ref{secpnjl0}, we generalize the PNJL0 model at $T=0$ by taking into account the dynamics of the strange~$s$ quark. The thermodynamics of the SU(3) model is studied for both cases, namely, symmetric quark matter, and charge neutral system of quarks and leptons in weak equilibrium. In the former, we verify a strong reduction of the constituent strange quark mass ($M_s$) at the chemical potential where the finite value of $\Phi$ emerges (deconfined phase). An application to hybrid stars is also performed. The results indicate that it is possible to generate hybrid stars compatible with the recent data from the Neutron Star Interior Composition Explorer (NICER) mission. Finally, summary and concluding remarks are presented in Sec.~\ref{secsummary}.

\section{Three-flavor Polyakov-Nambu-Jona-Lasinio model}
\label{secpnjl}

We start by presenting the Lagrangian density of the three-flavor version of the PNJL model, that includes a scalar and a vector 4-fermion interaction. It reads
\begin{align}
\mathcal{L}_{\mbox{\tiny PNJL}} &= \bar{q}(i\gamma_\mu D^\mu - \hat{m})q  - \mathcal{U}(\Phi,\Phi ^*,T)
\nonumber\\
&+\frac{G_s}{2}\sum_{a=0}^{8}\left[(\bar{q}\lambda_aq)^2-(\bar{q}\gamma_5 \lambda_aq)^2 \right]  
\nonumber\\
&-\frac{G_V}{2}\sum_{a=0}^{8}\left[(\bar{q}\gamma_\mu\lambda_aq)^2+(\bar{q}\gamma_\mu\gamma_5 \lambda_aq)^2 \right]
\nonumber\\
&+ K[\mbox{det}_f(\bar{q}(1-\gamma_5)q) + \mbox{det}_f(\bar{q}(1+\gamma_5)q)],
\label{dlpnjl}
\end{align}
where $q$ is a vector of three spinors $q_f$ for $f=u,d,s$, $\hat{m}=\mbox{diag}(m_u,m_d,m_s)$ is a matrix of current quark masses in flavor space, and $\lambda_a$ are the SU(3) Gell-Mann matrices. The last term, a determinant in flavor space, corresponds to the so called 't Hooft interaction~\cite{thooft} and is responsible for the large value of the $\eta'$ mass due to the anomaly $U_A(1)$~\cite{vogl}. Such a kind of interactions were firstly considered by Kobayashi and Maskawa~\cite{km}. The differences between this model and the NJL one~\cite{Nambu1,Nambu2,buballa,vogl2,outros,ric1,ric2,ric3} are the replacement of $\partial^\mu$ by $D^\mu\equiv\partial^\mu+iA^\mu $, where $A^\mu=\delta^\mu_0A_0$ 
and $A_0=gA_a^0\lambda_a / 2$ ($g$ is the gauge coupling), and the inclusion of the Polyakov potential $\mathcal{U}(\Phi,\Phi ^*,T)$. It depends on the traced Polyakov loop and its conjugate, $\Phi$ and $\Phi^*$, with $\Phi$ is defined in terms of $A_4=iA_0\equiv T\phi$ as
\begin{align}
\Phi&=\frac{1}{3}\rm{Tr}\left[\,\,\rm{exp}\left(i\int_0^{1/T}d\tau\,A_4\right)\right]
\nonumber \\
&=
\frac{1}{3}\left[\rm{e}^{i(\phi_3+\phi_8/\sqrt{3})}+\rm{e}^{i(-\phi_3+\phi_8/\sqrt{3})}
+\rm{e}^{-2i\phi_8/\sqrt{3}}\right],
\label{traced}
\end{align}
written in the Polyakov gauge where $\phi=\phi_3\lambda_3+\phi_8\lambda_8$. In the mean-field approximation, and using that $\Phi=\Phi^*$~\cite{epjc,mattos,weise4,weise6,bratovic,rossner2,Schramm,Dexheimer,Steinheimer}, Eq.~(\ref{dlpnjl}) becomes
\begin{align}
\mathcal{L}_{\mbox{\tiny PNJL}}^{\mbox{\tiny mfa}} &= \sum_f\bar{q}_f(i\gamma_\mu D^\mu - M_f)q_f
-G_s\sum_f{\rho_{sf}^2}
\nonumber\\
&+ G_V\sum_f\rho_f^2 - 4K\prod_f\rho_{sf} - \mathcal{U}(\Phi,T).
\label{dlpnjlmfa}
\end{align}

The grand canonical potential density obtained from Eq.~(\ref{dlpnjlmfa}) reads
\begin{align}
&\Omega_{\mbox{\tiny PNJL}} = G_s\sum_f\rho_{sf}^2 - G_V\sum_f\rho_f^2 + 4K\prod_f\rho_{sf} 
\nonumber \\
&- \frac{\gamma}{2\pi^2}\sum_f\int_0^{\Lambda}dk\,k^2(k^2+M_f^2)^{1/2}
\nonumber \\
&- \frac{\gamma}{6\pi^2}\sum_f\int_0^{\infty}\hspace{-0.3cm}\frac{dk\,k^4}{(k^2+M_f^2)^{1/2}}[f(E_f,T,\Phi)+\bar{f}(E_f,T,\Phi)]
\nonumber \\
&+\mathcal{U}(\Phi,T),
\label{omegapnjl}
\end{align}
with
\begin{eqnarray}
\rho_{sf} &=& \frac{\gamma}{2\pi^2}\int_0^{\infty}dk\,k^2\frac{M_f}{E_f}
\left[ f(E_f,T,\Phi) + \bar{f}(E_f,T,\Phi) \right] \nonumber\\
&-&\frac{\gamma}{2\pi^2}\int_0^{\Lambda}dk\,k^2\frac{M_f}{E_f}
\nonumber\\
&=&\left<\bar{q}_fq_f\right>,
\label{rhospnjl}
\end{eqnarray}
and
\begin{eqnarray}
\rho_f &=& -\frac{\partial \Omega_{\mbox{\tiny PNJL}}}{\partial\mu_f} 
\nonumber\\
&=& \frac{\gamma}{2\pi^2}\int_0^{\infty}dk\,k^2[f(E_f,T,\Phi) - \bar{f}(E_f,T,\Phi)]
\nonumber\\
&=&\left<\bar{q}_f\gamma_0q_f\right>
\label{rhopnjl}
\end{eqnarray}
Here $\rho_{sf}$ is the quark ``scalar'' density (quark condensate) related to flavor $f$, and $\rho_f$ is the ``vector'' quark density. Furthermore, $E_f=(k^2+M_f^2)^{1/2}$ and $\gamma=N_s\times N_c=6$ is the degeneracy factor given in terms of spin ($N_s=2$) and color ($N_c=3$) numbers. The constituent quark masses, $M_f$, are given in terms of the quark condensates as
\begin{eqnarray}
M_f = m_f -2G_s\rho _{sf}-2K\prod _{f'\neq f}\rho _{sf'}.
\label{mf}
\end{eqnarray}
As in the case of two-flavor PNJL model~\cite{newratti,FUKUSHIMA,weise1,weise2,rossner1,rossner2}, the functions $f(E_f,T,\Phi)$ and $\bar{f}(E_f,T,\Phi)$, given by
\begin{align}
&f(E_f,T,\Phi)=\nonumber\\
&\frac{\Phi e^{2(E_f-\tilde{\mu}_f)/T} + 2\Phi e^{(E_f-\tilde{\mu}_f)/T}+ 1}
{3\Phi e^{2(E_f-\tilde{\mu}_f)/T} + 3\Phi e^{(E_f-\tilde{\mu}_f)/T} + e^{3(E_f-\tilde{\mu}_f)/T} + 1}
\label{fdmp} 
\end{align}
and
\begin{align}
&\bar{f}(E_f,T,\Phi) =\nonumber\\
&\frac{ \Phi e^{2(E_f+\tilde{\mu}_f)/T}+2\Phi e^{(E_f+\tilde{\mu}_f)/T}+1}
{ 3\Phi e^{2(E_f+\tilde{\mu}_f)/T} + 3\Phi e^{(E_f+\tilde{\mu}_f)/T} + e^{3(E_f+\tilde{\mu}_f)/T} + 1},
\label{fdmap}
\end{align}
are the generalized Fermi-Dirac distributions for quarks and antiquarks with $\tilde{\mu}_f=\mu_f - 2G_V\rho_f$. It is worth to notice that the vector interaction~\cite{bratovic,dynamical,Vector-Int} produces a shift in the chemical potentials ($\mu_f$), exactly as it occurs in the NJL model. Such distributions are one of two basic differences between PNJL and NJL models. The other one is the Polyakov potential $\mathcal{U}(\Phi,T)$, a quantity that determines the  thermodynamics of the gluonic pure gauge sector. It presents some forms such as the following ones
\begin{align}
\frac{\mathcal{U}_{\mbox{\tiny RTW05}}}{T^4} &= -\frac{b_2(T)}{2}\Phi^2
- \frac{b_3}{3}\Phi^3 + \frac{b_4}{4}\Phi^4,  
\label{rtw05} \\
\frac{\mathcal{U}_{\mbox{\tiny RRW06}}}{T^4} &= -\frac{b_2(T)}{2}\Phi^2
+ b_4(T)\mbox{ln}(1 - 6\Phi^2 + 8\Phi^3 - 3\Phi^4), 
\label{rrw06} \\
\frac{\mathcal{U}_{\mbox{\tiny FUKU08}}}{b\,T} &= -54e^{-a/T}\Phi^2 
- \mbox{ln}(1 - 6\Phi^2 + 8\Phi^3 - 3\Phi^4), 
\label{fuku08} \\
\mathcal{U}_{\mbox{\tiny DS10}} &= (a_0T^4 + a_1\mu^4 + a_2T^2\mu^2)\Phi^2 + \mathcal{U}_0(\Phi),
\label{ds10}
\end{align}
with
\begin{eqnarray}
b_2(T) &=& a_0 + a_1\left(\frac{T_0}{T}\right) + a_2\left(\frac{T_0}{T}\right)^2 
+ a_3\left(\frac{T_0}{T}\right)^3,
\label{b2t}\\
b_4(T) &=& b_4\left(\frac{T_0}{T}\right)^3,
\end{eqnarray}
and
\begin{eqnarray}
\mathcal{U}_0(\Phi)\equiv a_3T_0^4\mbox{ln}(1-6\Phi^2+8\Phi^3-3\Phi^4),
\label{u0}
\end{eqnarray}
with $a$, $b$, $a_0$, $a_1$, $a_2$, $a_3$, $b_3$ and $b_4$ being dimensionless free parameters. $T_0$ is defined as the transition temperature for the pure gauge system. These potentials are named as RTW05~\cite{weise1}, RRW06~\cite{weise2,weise4}, FUKU08~\cite{fukushima3} and DS10~\cite{Schramm,Dexheimer}.  The traced Polyakov loop is found from $\partial\Omega_{\mbox{\tiny PNJL}}/\partial\Phi=0$, and the current quark masses $M_{u,d,s}$ are determined through solutions of Eqs.~(\ref{mf}), in which condensates ${\rho_s}_{u,d,s}$ given in Eqs.~(\ref{rhospnjl}) are also used.

Finally, from Eq.~(\ref{omegapnjl}) all the other thermodynamical quantities can be determined, namely, pressure, entropy  and energy densities. The expressions are obtained from $P_{\mbox{\tiny PNJL}} = -\Omega_{\mbox{\tiny PNJL}}$, $\mathcal{S}_{\mbox{\tiny PNJL}}=-\partial\Omega_{\mbox{\tiny PNJL}}/\partial T$, and $\mathcal{E}_{\mbox{\tiny PNJL}} = T\mathcal{S}_{\mbox{\tiny PNJL}} - P_{\mbox{\tiny PNJL}} + \sum_f\mu_f\rho_f$, respectively.

\section{PNJL0 model: three-flavor case}
\label{secpnjl0}

\subsection{Formulation}

At this point, we remind the reader that the  the PNJL model equations, previously presented, are reduced to those related to the NJL one at zero temperature regime. The origin of this result is twofold. First, one has that the generalized Fermi-Dirac distributions, given in Eqs.~(\ref{fdmp}) and~(\ref{fdmap}), become step functions $\theta(k_{Ff}-k)$ at $T=0$ ($k_{Ff}$ is the Fermi momentum of the quark $f$). Therefore, $\Phi$ disappears in all momentum integrals. The second reason is due to the gluonic contribution enclosed by the Polyakov potential. Notice that $\mathcal{U}(\Phi,0)$ vanishes for the most known versions. In this case, Eq.~(\ref{omegapnjl}) reads
\begin{align}
&\Omega_{\mbox{\tiny PNJL}}(T=0) = G_s\sum _f \rho _{sf}^2 -G_V\sum _f \rho _f^2 + 4K\prod _f \rho _{sf} 
\nonumber\\
&- \frac{\gamma}{2\pi ^2}\sum _f\int_0^\Lambda dk\,k^2(k^2+M_f^2)^{1/2}\nonumber \\
&- \frac{\gamma}{6\pi ^2}\sum _f\int_0^{k_{Ff}}\frac{dk\,k^4}{(k^2+M_f^2)^{1/2}} 
= \Omega_{\mbox{\tiny NJL}}(T=0).
\label{omega0}
\end{align}

In order to circumvent this problem, we implemented, in Ref.~\cite{epjc}, the introduction of $\Phi$ in the SU(2) NJL model at $T=0$ by requiring that the strengths of the scalar and vector channels are vanishing in the deconfined phase, i.e., at $\Phi=1$. In that approach we incorporate expected effects from QCD that at low densities, and corresponding large inter-particle distances, quarks should interact strongly while at short distance the interaction should be weakened. The former is associated to the nonperturbative infrared physics from QCD that enhances the interaction between the effective degrees of freedom, related to the quarks in the model. The latter should represent the ultraviolet physics of perturbative QCD where the quarks interact weakly due to the asymptotic freedom phenomenon.  Therefore, this physics is incorporated in a dynamical way such that at low densities quarks interact strongly and at large densities weakly towards to the deconfinement phase transition. For this purpose, we use the traced Polyakov loop as an effective scalar background field at the zero temperature regime. Although this quantity is not strictly defined at $T=0$, we use it to bring the complexity of QCD dynamics by incorporating, in a effective way, the transition from strong (infrared) to weak (ultraviolet) regimes of the quark-quark interaction. Here, we extended this phenomenology also to the three-flavor version of the model. The modifications in the couplings for this case are 
\begin{eqnarray}
G_s \rightarrow \mathcal{G}_s(G_s,\Phi) &=& G_s( 1-\Phi^2 ),
\label{gsphi}
\end{eqnarray}
\begin{eqnarray}
G_V \rightarrow \mathcal{G}_V(G_V,\Phi) &=& G_V( 1-\Phi^2 ),
\label{gvphi}
\end{eqnarray}
and 
\begin{eqnarray}
K \rightarrow  \mathcal{K}(K,\Phi) &=& K( 1-\Phi^2 ).
\label{kphi}
\end{eqnarray}

Such changes lead to 
\begin{align}
\Omega_{\mbox{\tiny PNJL0}} &= G_s\sum _f \rho _{sf}^2 -G_V\sum _f \rho _f^2 + 4K\prod _f \rho _{sf} 
\nonumber\\
&-\frac{\gamma}{2\pi ^2}\sum _f\int_0^\Lambda dk\,k^2(k^2+M_f^2)^{1/2}
\nonumber \\
&-\frac{\gamma}{6\pi ^2}\sum_f\int_0^{k_{Ff}}\frac{dk\,k^4}{(k^2+M_f^2)^{1/2}}
+\mathcal{U}(\rho_f,\rho_{sf},\Phi)
\label{ompnjl0}
\end{align}
with
\begin{align}
&\mathcal{U}(\rho_f,\rho_{sf},\Phi)\equiv\mathcal{U}(\rho_u,\rho_d,\rho_s,\rho_{su},\rho_{sd},\rho_{ss},\Phi) =
\nonumber\\
&=G_V\Phi^2\sum _f \rho _f^2
-G_s\Phi^2\sum _f \rho _{sf}^2 
-4K\Phi^2\prod _f \rho _{sf}
\nonumber\\
&+ \mathcal{U}_0(\Phi).
\label{upnjl0}
\end{align}
By choosing to incorporate the terms $G_s\Phi^2$, $G_V\Phi^2$ and $K\Phi^2$ of the grand canonical potential into $\mathcal{U}$ defined in Eq.~(\ref{upnjl0}), it is possible to identify Eq.~(\ref{ompnjl0}) as the zero temperature version of Eq.~(\ref{omegapnjl}). As in Ref.~\cite{epjc}, we refer to this construction as the PNJL0 model, now written in its SU(3) version. Quark masses and chemical potentials are given, respectively, by
\begin{eqnarray}
M_f = m_f -2G_s(1-\Phi^2)\rho _{sf}-2K(1-\Phi^2)\prod _{f'\neq f}\rho _{sf'},\qquad
\label{mfpnjl0}
\end{eqnarray}
and
\begin{align}
\mu _f = (k_{Ff}^2 + M_f^2)^{1/2}+ 2G_V(1-\Phi^2)\rho _f.
\end{align}
where 
\begin{eqnarray}
\rho_f=\frac{\gamma}{6\pi^2}k_{Ff}^3
\end{eqnarray}
and
\begin{eqnarray}
\rho_{sf}=-\frac{\gamma M_f}{2\pi^2}\int_{k_{Ff}}^\Lambda \frac{dk\,k^2}{(k^2+M_f^2)^{1/2}}.
\label{rhospnjl0}
\end{eqnarray}
Notice that in the three-flavor version of the PNJL0 model, the modification of the couplings, Eqs.~(\ref{gsphi})-(\ref{kphi}), also induces the definition of a new Polyakov potential, Eq.~(\ref{upnjl0}), in which the backreaction of quarks in the gluonic sector happens, as well as the inverse one, namely, gluons affecting quarks. This last effect is intrinsic in the original PNJL model but the former is absent. In the SU(3) PNJL0 model, the backreaction is complete, i.e., each sector interacts each other, exactly as in its two-flavor version~\cite{epjc}. We remark that the modifications pointed out in Eqs.~(\ref{gsphi})-(\ref{kphi}) lead to the Polyakov potential $\mathcal{U}=G_V\Phi^2\sum _f \rho _f^2-G_s\Phi^2\sum _f \rho _{sf}^2 -4K\Phi^2\prod _f \rho _{sf}$. However, this potential is not able to generate nonvanishing solutions for $\Phi$ when the condition $\partial\Omega_{\mbox{\tiny PNJL0}}/\partial\Phi=0$ is applied. Therefore, we add the term $\mathcal{U}_0$ given in Eq.~(\ref{u0}) in the definition of $\mathcal{U}$ and verify that it is responsible to ensure $\Phi\ne0$. Moreover, it is also important in order to restrict the traced Polyakov loop in the range of $0\leqslant\Phi\leqslant 1$~\cite{Schramm,Dexheimer}. Finally, the pressure and energy density of the SU(3) PNJL0 model, obtained from Eq.~(\ref{ompnjl0}), are written as
\begin{align}
&P_{\mbox{\tiny PNJL0}} = -G_s\sum _f \rho _{sf}^2 + G_V\sum _f \rho _f^2 - 4K\prod _f \rho _{sf} \nonumber\\
&+\frac{\gamma}{2\pi ^2}\sum _f\int_0^\Lambda dk\,k^2(k^2+M_f^2)^{1/2} 
\nonumber \\
&+\frac{\gamma}{6\pi ^2}\sum _f\int_0^{k_{Ff}}\frac{dk\,k^4}{(k^2+M_f^2)^{1/2}} 
\nonumber\\
&-\mathcal{U}(\rho_f,\rho_{sf},\Phi) +\Omega_{\mbox{\tiny vac}}
\end{align}
and
\begin{align}
&\mathcal{E}_{\mbox{\tiny PNJL0}} = G_s\sum _f\rho _{sf}^2 + G_V \sum _f\rho _{f}^2
+ 4K\prod _f \rho _{sf} 
\nonumber\\
&-\frac{\gamma}{2\pi^2} \sum _f\int_{k_{Ff}}^\Lambda dk\,k^2(k^2+M_f^2)^{1/2}
-2G_V\Phi^2\sum _f \rho _{f}^2 
\nonumber\\
&  +\mathcal{U}(\rho_f,\rho_{sf},\Phi) - \Omega_{\mbox{\tiny vac}},
\end{align}
respectively. The constant $\Omega_{\mbox{\tiny vac}}=-P_{\mbox{\tiny vac}}$ is added to the pressure and energy density in order to ensure that in vacuum ($\rho_u=\rho_d=\rho_s=0$) one has $\Omega_{\mbox{\tiny PNJL0}}(\rho_f=0)=-P_{\mbox{\tiny PNJL0}}(\rho_f=0)=0$ and $\mathcal{E}_{\mbox{\tiny PNJL0}}(\rho_f=0)=0$.

\subsection{Thermodynamics of the model (symmetric quark matter)}

We remark that from now on, all results are obtained at zero temperature regime.

Since the main equations of the model are defined, we are able to explore some thermodynamical features of the SU(3) PNJL0 model. Firstly, we need to define values for its free parameters, namely, $G_s$, $G_V$, $K$, $m_u$, $m_d$, $m_s$ and $\Lambda$. Here we use the RKH parametrization~\cite{buballa,rkh} in which $G_s=3.67/\Lambda^2$, $K=-12.36/\Lambda^5$, $m_u=m_d=5.5$~MeV, $m_s=140.7$~MeV, and $\Lambda=602.3$~MeV. The last quantity is the cutoff parameter. As in the NJL model, the PNJL0 one is nonrenormalizable. Therefore, it is needed to adopt a certain regularization scheme in order to avoid divergent contributions in the momentum integrals. Here we use a three-momentum cutoff for this purpose~\cite{buballa,greiner,leupold}. This parameters set produces the values of $f_\pi=92.4$~MeV, $m_\pi=135$~MeV, $m_K=497.7$~MeV, $m_\eta=514.8$~MeV, and $m_\eta'=957.8$~MeV, respectively, for the pion decay constant and mass, and for the masses of pseudoscalar mesons $K$, $\eta$, and $\eta'$~\cite{buballa,rkh}. For the vacuum, it also leads to $M_u^{\mbox{\tiny (vac)}}=M_d^{\mbox{\tiny (vac)}}=367.6$~MeV, $M_{s}^{\mbox{\tiny (vac)}}=549.5$~MeV,  $\langle \overline {u}u \rangle^{1/3}_{\mbox{\tiny (vac)}}=\langle \overline {d}d \rangle^{1/3}_{\mbox{\tiny (vac)}}=-241.9$~MeV, and $\langle \overline {s}s \rangle^{1/3}_{\mbox{\tiny (vac)}}=-257.7$~MeV. 

With regard to the strength of the vector channel, we remind that if a Fierz transformation of the color-current interaction is performed, the value of $G_V=0.5G_s$ is found~\cite{buballa,vogl2}. In principle, this ``canonical'' value should be used in all calculations involving NJL/PNJL models with the vector channel included. However, $G_V$ is often taken as a free parameter. Some authors use this freedom to fix the vector meson masses, such as the~$\rho$ meson~\cite{lutz}. Other ones also point out to changes in the magnitude of $G_V$ due to medium effects. For instance, instanton-anti-instanton pairing effects at high temperature can change its magnitude~\cite{shuryak}. Furthermore, it is also known that~$G_V$ plays a significant hole in the phase diagram of the strongly interacting matter since it shifts the location of the critical end point (CEP)~\cite{bratovic,kashiwa}. The increasing of $G_V$ induces the disappearance of the CEP and, consequently, the vanishing of the first order phase transition line. Therefore, as in Ref.~\cite{epjc}, we adopt here the approach of setting $G_V$ as a free parameter in order to verify its influence in our calculations.

Finally, the gluonic sector of the model present two more constants, namely, $T_0$ and~$a_3$, appearing in the last term of the Polyakov potential, Eq.~(\ref{upnjl0}). The former is fixed to $190$~MeV~\cite{epjc,ratti1}. Concerning the constant $a_3$, we remind that it appears in the quantity denoted by $\mathcal{U}_0$, Eq.~(\ref{u0}), firstly used in Ref.~\cite{Schramm}. In those works, the authors remark that the free parameters presented in the Polyakov Potential~$\mathcal{U}$, that contains~$\mathcal{U}_0$, are determined in order to reproduce lattice data, as well as known information about the phase diagram. In this case, the value obtained by them is $a_3=-0.4$. However, it is worth to notice that their model is completely different from the PNJL0 model considered here. If we take $a_3=-0.4$, no solutions of $\Phi\neq 0$ are found, as we will show in the first figure later on. Therefore, one has $G_V$ and~$a_3$ as free parameters of the SU(3) PNJL0 model, as well as they are in its two-flavor version~\cite{epjc}.

At zero temperature, one can also use the thermodynamical relation given by
\begin{align}
E + PV &= \mu_BN_B + \mu_SN_S + \mu_IN_I 
\nonumber\\
&= \mu_uN_u + \mu_dN_d + \mu_sN_s, 
\end{align}
in order to write the quark chemical potentials as
\begin{align}
\mu_u = \frac{\mu_B}{3} + \frac{\mu_I}{2},\,\,\,
\mu_d = \frac{\mu_B}{3} - \frac{\mu_I}{2},\,\,\,
\mu_s = \frac{\mu_B}{3} - \mu_S,
\label{cp}
\end{align}
i.e., in terms of the baryonic, isospin, and strangeness chemical potentials, namely, $\mu_B$, $\mu_I$, and $\mu_S$, respectively.  Firstly, we work at the so called symmetric quark matter, the one in which all quark potentials are the same, namely, 
\begin{align}
\mu_u=\mu_d=\mu_s\equiv\mu,
\label{symmetric}
\end{align}
situation found by taking $\mu_I=\mu_S=0$ in Eq.~(\ref{cp}). At this point, we remark to the reader that the concept of symmetric matter in this context is different from the one used in nuclear matter described by hadronic models~\cite{had1}, in which nucleons are in equal number and present identical chemical potentials, since they have the same rest masses and densities. In the case of the three-flavor quark system studied here, $m_u=m_d$ but $m_s$ is quite different. Therefore, densities of the light quarks are equal each other but different from the strange quark one, despite the validity of Eq.~(\ref{symmetric}). 

\begin{figure}[!htb] 
\centering
\includegraphics[scale=0.337]{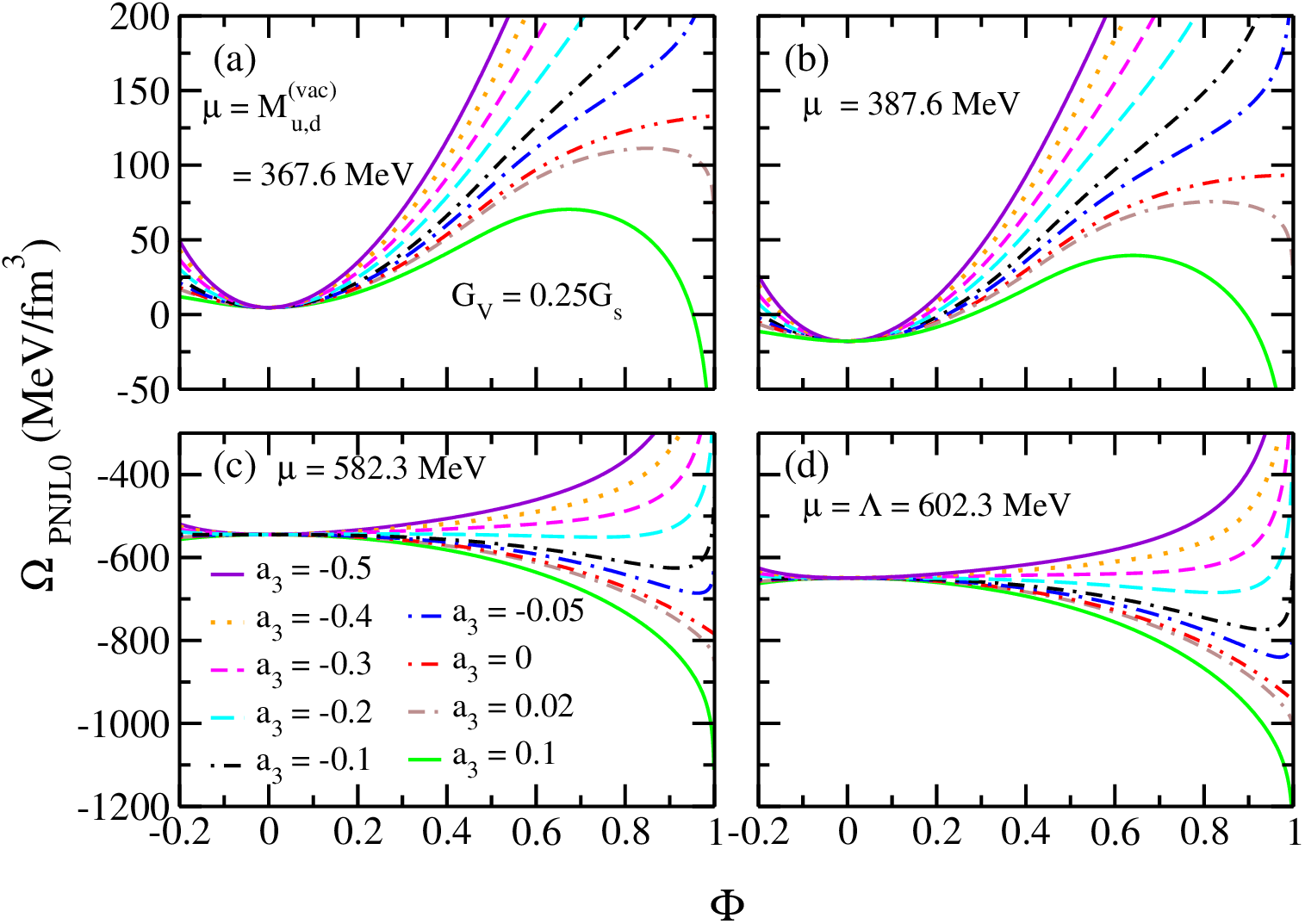}
\includegraphics[scale=0.337]{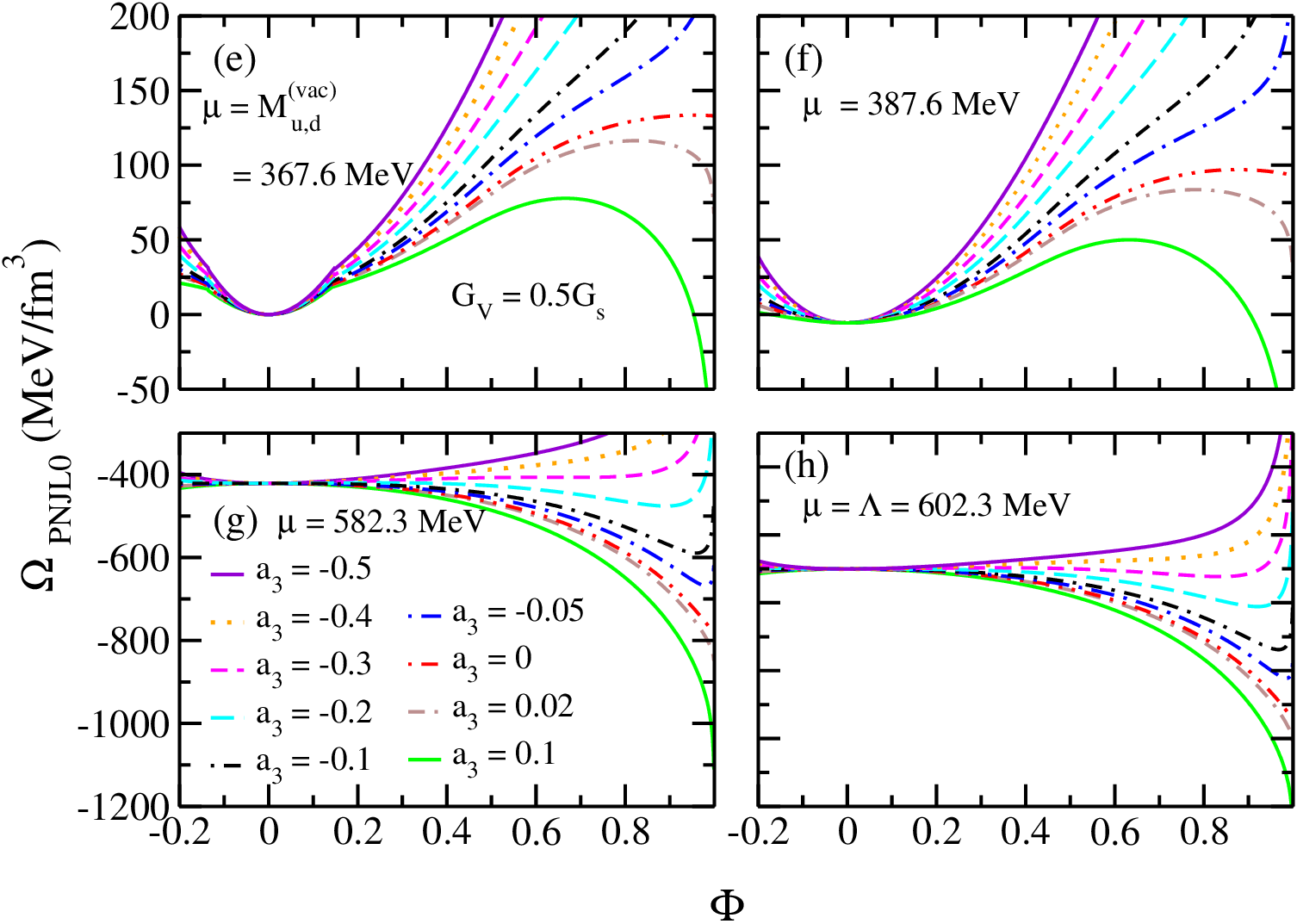}
\caption{ $\Omega_{\mbox{\tiny PNJL0}}$ as a function of $\Phi$ for $G_V=0.25G_s$ (a,b,c,d) and $G_V=0.5G_s$ (e,f,g,h) for different~$a_3$. The common quark chemical potential is given by (a and b) $\mu=M_u^{\mbox{\tiny (vac)}}=M_d^{\mbox{\tiny (vac)}}=367.6$~MeV, (b and f)~$\mu=387.6$~MeV, (c and g)~$\mu=582.3$~MeV, and (d and h)~$\mu=\Lambda=602.3$~MeV.}
\label{omphi}
\end{figure}

As a first investigation of the symmetric quark matter described by the SU(3) PNJL0 model, we show in Fig.~\ref{omphi} $\Omega_{\mbox{\tiny PNJL0}}$ as a function of the traced Polyakov loop for $G_V=0.25G_s$ and for the canonical relation $G_V=0.5G_s$ found through Fierz transformation of the current-current interaction~\cite{buballa,vogl2}.
In order to construct these curves, we use Eq.~(\ref{ompnjl0}) with the constituent quark masses found by the transcendental equations given by Eqs.~(\ref{mfpnjl0}) and~(\ref{rhospnjl0}). $\Phi$ is free to run in this case. Our analysis points out that in the range of chemical potentials from $\mu=M_{u,d}^{\mbox{\tiny (vac)}}$ to $\mu=\Lambda$, it is possible to find curves in which $\Omega_{\mbox{\tiny PNJL0}}$ present global minima for nonvanishing values of the traced Polyakov loop, see Figs.~\ref{omphi}c and~\ref{omphi}d, for instance. For the curve presenting $a_3=-0.2$ and  $G_V=0.25G_s$, we clearly see a single minimum of $\Omega_{\mbox{\tiny PNJL0}}$ at $\Phi=0$ for $\mu=M_{u,d}^{\mbox{\tiny (vac)}}$ (panel a). However, $\Phi\sim0.82$ produces a minimum of $\Omega_{\mbox{\tiny PNJL0}}$ for $\mu=\Lambda$ (panel d). As in the SU(2) case~\cite{epjc}, we verify a confinement/deconfinement phase transition in the model, i.e, the existence of null and nonvanishing values for the order parameter $\Phi$ as a function of the chemical potential. One can observe in panels e, f, g and h that a similar behavior is verified for the curves in which $G_V=0.5G_s$. Another feature observed here, and also present in the SU(2) version of the model~\cite{epjc}, is the need of $G_V\neq 0$ values in order to become possible the existence of a global minimum of $\Omega_{\mbox{\tiny PNJL0}}$ at $\Phi\neq 0$. For $G_V=0$, the only minimum is found at $\Phi=0$ for different $a_3$ values (figure not shown). The vector interaction still plays an important role even in the SU(3) version of the PNJL0 model.

\begin{figure}[!htb] 
\centering
\includegraphics[scale=0.33]{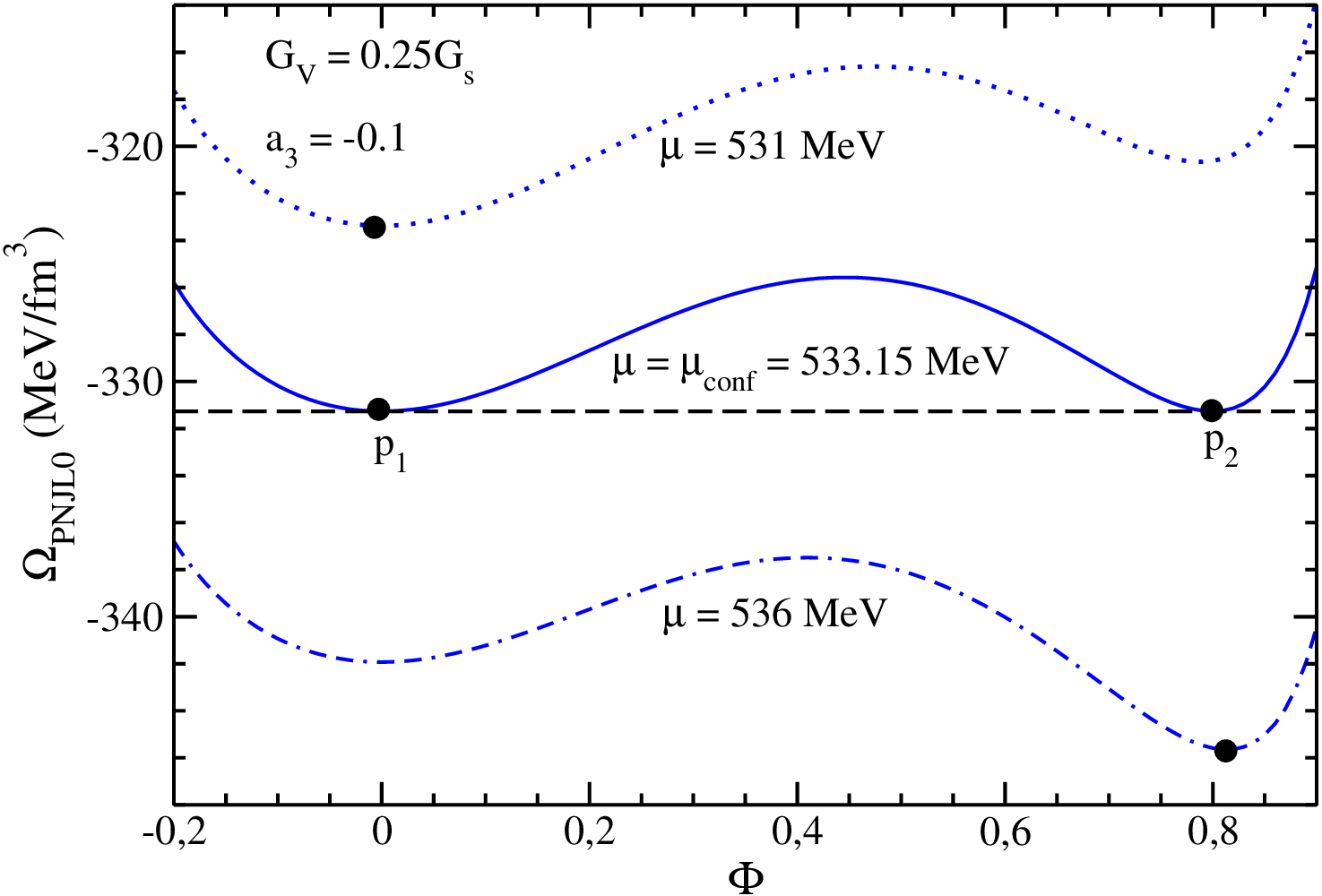}
\caption{$\Omega_{\mbox{\tiny PNJL0}}$ as a function of $\Phi$ for $G_V=0.25G_s$ and $a_3=-0.1$ for three different values of $\mu$.} 
\label{omphiconf}
\end{figure}

\begin{figure}[!htb] 
\centering
\includegraphics[scale=0.33]{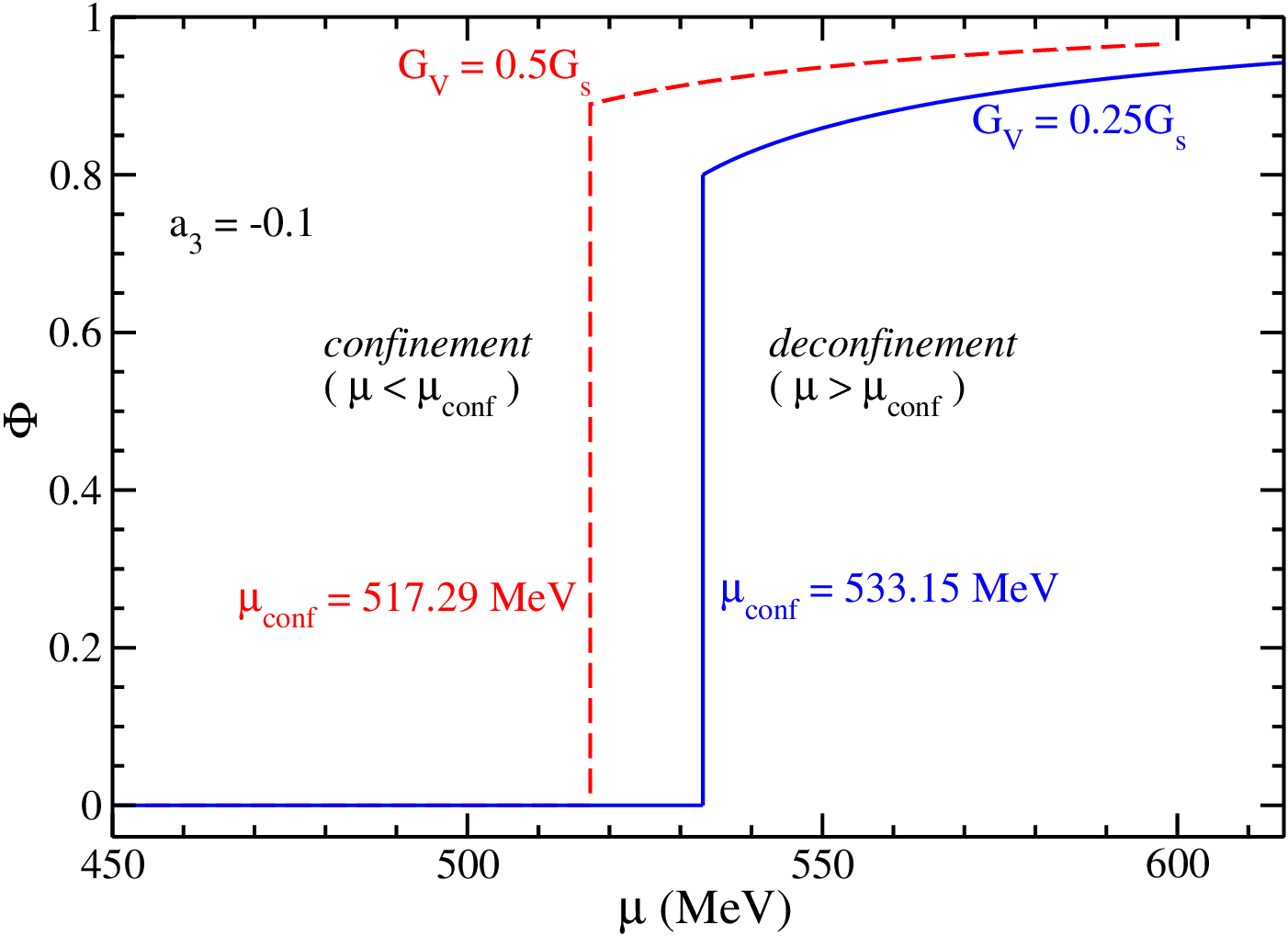}
\caption{Traced Polyakov loop as a function of the common quark chemical potential for $G_V=0.25G_s$ and $G_V=0.5G_s$, both for $a_3=-0.1$. These parametrizations lead to $\mu_{\mbox{\tiny conf}}=533.15$~MeV and $\mu_{\mbox{\tiny conf}}=517.29$~MeV, respectively.} 
\label{phimu}
\end{figure}

The analysis of the $\Omega_{\mbox{\tiny PNJL0}}\times\Phi$ curves is also useful in order to determine the chemical potential in which the confinement/deconfinement phase transition takes place, named here as $\mu_{\mbox{\tiny conf}}$. For the model with $G_V=0.25G_s$ and $a_3=-0.1$, for example, we see from Fig.~\ref{omphiconf} that $\mu_{\mbox{\tiny conf}}=533.15$~MeV.
This particular chemical potential value is defined as the one that generates a grand canonical potential presenting two minima with the same value for  $\Omega_{\mbox{\tiny PNJL0}}$, i. e., $\Omega_{\mbox{\tiny PNJL0}}(\Phi_1)=\Omega_{\mbox{\tiny PNJL0}}(\Phi_2)$. In that case, $\Phi_1$ and $\Phi_2$ delimit the boundaries of the thermodynamical phases associated to quark confinement and deconfinement. The traced Polyakov loop is the order parameter of this transition. The procedure described above can be used to construct the $\Phi\times\mu$ curve depicted in Fig.~\ref{phimu}, in this case for $G_V=0.25G_s$ and $G_V=0.5G_s$.
From this figure, we notice the quark confinement region determined by the condition given by $\mu<\mu_{\mbox{\tiny conf}}$. On the other hand, deconfinement is achieved through a first order phase transition, in the region where $\mu>\mu_{\mbox{\tiny conf}}$, according to the thermodynamics presented by the model. This feature was also observed in Ref.~\cite{Schramm} in which the authors included the traced Polyakov loop in a hadronic SU(3) nonlinear $\sigma$ model along with quark degrees of freedom. Furthermore, in Ref.~\cite{pnjl0outro} another version of a three-flavor PNJL model at zero temperature was studied. Specifically, a  dependence on the quark density was introduced in the $b_2(T)$ function of the Polyakov potential given in Eq.~(\ref{rrw06}). In that model, a first order phase transition is also verified in the $\Phi(\mu)$ curve. Another effect verified in Fig.~\ref{phimu} is the decreasing of $\mu_{\mbox{\tiny conf}}$ as $G_V$ increases, exactly as in the SU(2) case~\cite{epjc}.

\begin{figure}[!htb] 
\centering
\includegraphics[scale=0.33]{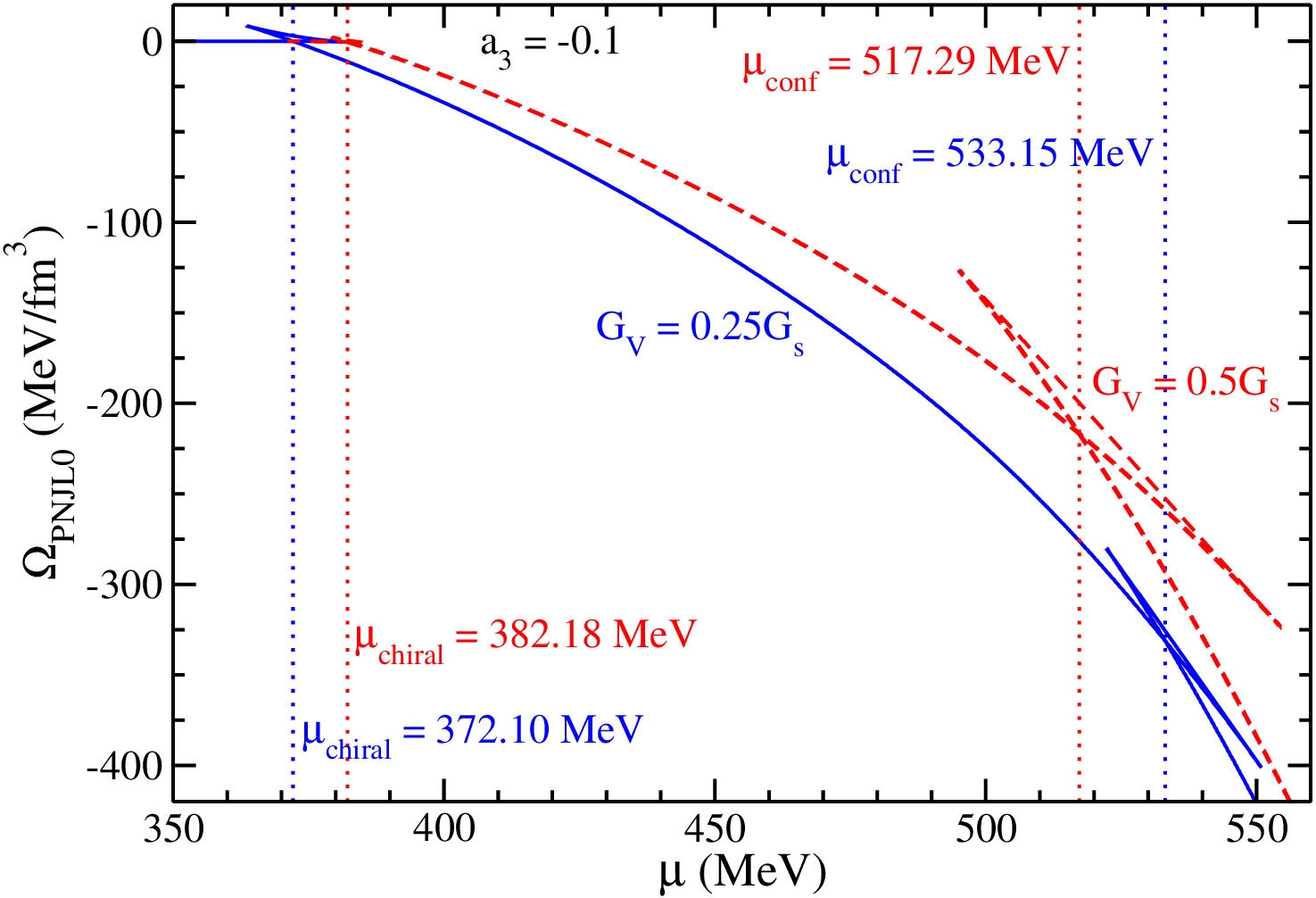}
\caption{$\Omega_{\mbox{\tiny PNJL0}}$ as a function of the common quark chemical potential for $G_V=0.25G_s$ and $G_V=0.5G_s$, both for $a_3=-0.1$.} 
\label{omegamu}
\end{figure}

Now we evaluate three-flavor PNJL0 model equations by solving Eqs.~(\ref{mfpnjl0}) and~(\ref{rhospnjl0}) together with condition $\partial\Omega_{\mbox{\tiny PNJL0}}/\partial\Phi=0$. In the SU(3) case, this procedure consists in searching for solutions of $M_u=M_d$, $M_s$ and $\Phi$, obtained from the set of 3 coupled equations aforementioned. By following this method, we are able to evaluate Eq.~(\ref{ompnjl0}), with result displayed in Fig.~\ref{omegamu}.
One can notice a typical signal of a system presenting first order phase transition~\cite{callen,yazaki,theis}, since Fig.~\ref{omegamu} shows that $\Omega_{\mbox{\tiny PNJL0}}$ (thermodynamical potential that describes the system) is multi-valued. A requirement from thermodynamical stability~\cite{callen} imposes that branches associated to unstable and metastable solutions is removed from the final curve. The ``crossing points'' indicate the chemical potential values related to the first order phase transitions. In the case of the PNJL0 model, there are two of them, one from confinement/deconfinement and another one from restored/broken chiral symmetry of the light quarks, that give rise to $\mu_{\mbox{\tiny conf}}=533.15$~MeV and $\mu_{\mbox{\tiny chiral}}=372.10$~MeV, respectively, for the parametrization of the model in which $G_V=0.25G_s$ and $a_3=-0.1$. Notice that at $\mu=\mu_{\mbox{\tiny conf}}$, we find the same value for $\Omega_{\mbox{\tiny PNJL0}}$  related to the two global minima of Fig.~\ref{omphiconf}, namely, $\Omega_{\mbox{\tiny PNJL0}}(\mu_{\mbox{\tiny conf}})\sim -331$~MeV/fm$^3$. This feature confirms the equivalence of the two methods used to define the first order phase transition points in such systems. It is also worth to notice the decrease of $\mu_{\mbox{\tiny conf}}$ as $G_V$ increases, exactly as pointed out in Fig.~\ref{phimu}.

\begin{figure}[!htb] 
\centering
\includegraphics[scale=0.33]{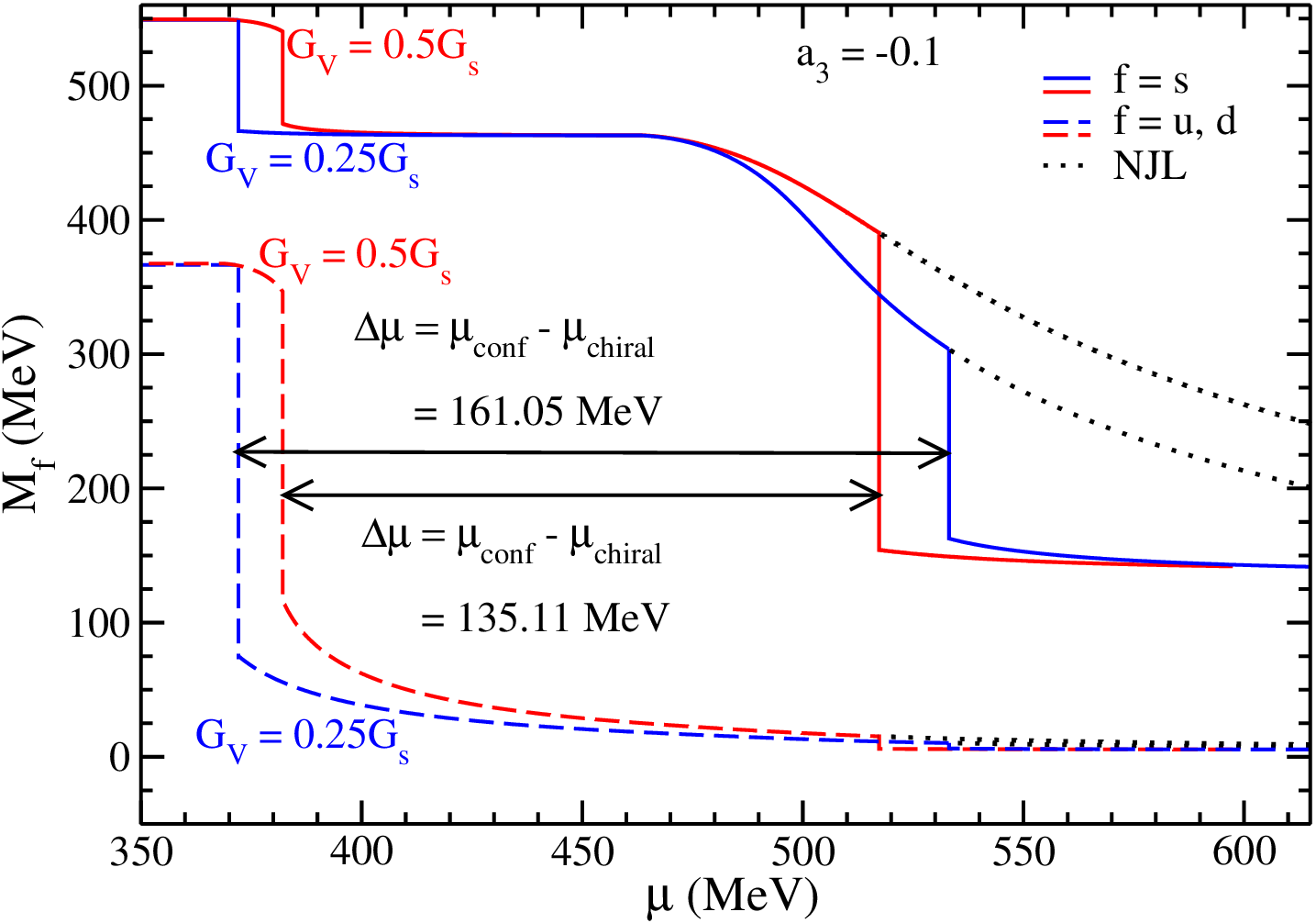}
\caption{Full and dashed lines: constituent quark masses as a function of the common quark chemical potential for $G_V=0.25G_s$ and $G_V=0.5G_s$, both for $a_3=-0.1$ (PNJL0 model). Dotted lines: SU(3) NJL model.} 
\label{masses}
\end{figure}

In the region where $\mu_{\mbox{\tiny chiral}}<\mu<\mu_{\mbox{\tiny conf}}$ we can recognize, as in the SU(2) version of the model~\cite{epjc}, a particular thermodynamical phase in which the light quarks present a very low mass in comparison to their vacuum values (almost fully restored chiral symmetry), and with $\Phi=0$ (confined quarks), i.e, the so called quarkyonic phase, pointed out in Refs.~\cite{fukushima3,abuki,mcnpa09,buisseret,mcnpa07,mcnpa08}, for example.  The quarkyonic phase related to the light quark sector can also be identified by looking at the constituent quark masses as a function of $\mu$, as displayed in Fig.~\ref{masses}. 
One can notice two sharp phase transitions occurring at $\mu_{\mbox{\tiny chiral}}=372.10$~MeV and $\mu_{\mbox{\tiny conf}}=533.15$~MeV, for $G_V=0.25G_s$, with the difference defining the ``size'' of the phase, in this case, $\Delta\mu=\mu_{\mbox{\tiny conf}} - \mu_{\mbox{\tiny chiral}}=161.05$~MeV. As we can also see, this difference is decreased for $G_V=0.5G_s$. More specifically, for the SU(3) PNJL0 model $\Delta\mu$ depends on both, the strength of the vector channel, $G_V$, and the $a_3$ parameter related to the Polyakov potential. In Tables~\ref{taba3} and ~\ref{tabgv} we present some values of $\Delta\mu$ by running $a_3$ and $G_V$ independently.
\begin{table}[!htb]
\caption{Parameter $a_3$ of the SU(3) PNJL0 model along with $\mu_{\mbox{\tiny conf}}$ and $\Delta\mu$. For each line one has $G_V=0.25G_s$ and $\mu_{\mbox{\tiny chiral}}=372.10$~MeV.}
\centering
\begin{tabular}{ccc}
\hline
$a_3$ & $\mu_{\mbox{\tiny conf}}$ (MeV)  & $\Delta\mu$ (MeV) \\
\hline
\hline
$-0.27$ & $602.30$ & $230.20$ \\
$-0.20$ & $576.74$ & $204.64$ \\ 
$-0.15$ & $555.87$ & $183.77$ \\ 
$-0.10$ & $533.15$ & $161.05$ \\
$-0.08$ & $523.40$ & $151.30$ \\
$-0.06$ & $513.14$ & $141.04$ \\
$-0.05$ & $507.75$ & $135.65$ \\
\hline
\label{taba3}
\end{tabular}
\end{table}

\begin{table}[!htb]
\caption{Parameter $G_V$ of the SU(3) PNJL0 model along with $\mu_{\mbox{\tiny conf}}$, $\mu_{\mbox{\tiny chiral}}$ and $\Delta\mu$. For each line we fixed $a_3 = -0.1$.}
\centering
\begin{tabular}{cccc}
\hline
\hline
$G_V/G_s$ & $\mu_{\mbox{\tiny conf}}$ (MeV) & $\mu_{\mbox{\tiny chiral}}$ (MeV) & $\Delta\mu$ (MeV) \\
\hline
\hline
$0.50$  & $517.29$ & $382.18$ & $135.11$ \\ 
$0.45$  & $519.38$ & $380.21$ & $139.17$ \\ 
$0.40$  & $521.82$ & $378.22$ & $143.60$ \\ 
$0.35$  & $524.75$ & $376.21$ & $148.54$ \\
$0.30$  & $528.41$ & $374.18$ & $154.23$ \\
$0.25$  & $533.15$ & $372.10$ & $161.05$ \\ 
$0.20$  & $539.70$ & $370.00$ & $169.70$ \\ 
\hline
\label{tabgv}
\end{tabular}
\end{table}

From Fig.~\ref{masses}, we can also notice a strong reduction of the constituent strange quark mass at $\mu=\mu_{\mbox{\tiny conf}}$.  At this point, $M_s$ undergoes a reduction of almost $50\%$ for $G_V=0.25G_s$, while for $G_V=0.5G_s$ the reduction is somewhat larger. The origin of this feature is the emergence of~$\Phi$ exactly at this chemical potential, see Fig.~\ref{phimu}. The decreasing of the couplings $\mathcal{G}_s(\Phi)$, $\mathcal{G}_V(\Phi)$ and $\mathcal{K}(\Phi)$ when $\Phi$ becomes nonvanishing induces a reduction in the constituent quark masses, including the strange quark one. We reinforce that such an effect is not present in the original SU(3) NJL model, also show in Fig.~\ref{masses} (in that case changes in $M_s$ are only gradual, see the dotted line). For the light sector, on the other hand, it is also verified a decreasing of $M_{u,d}$ in the PNJL0 model, however, with much smaller intensity, since at $\mu=\mu_{\mbox{\tiny conf}}$ the light quark masses are nearly vanishing already.
\begin{figure}[!htb] 
\centering
\includegraphics[scale=0.33]{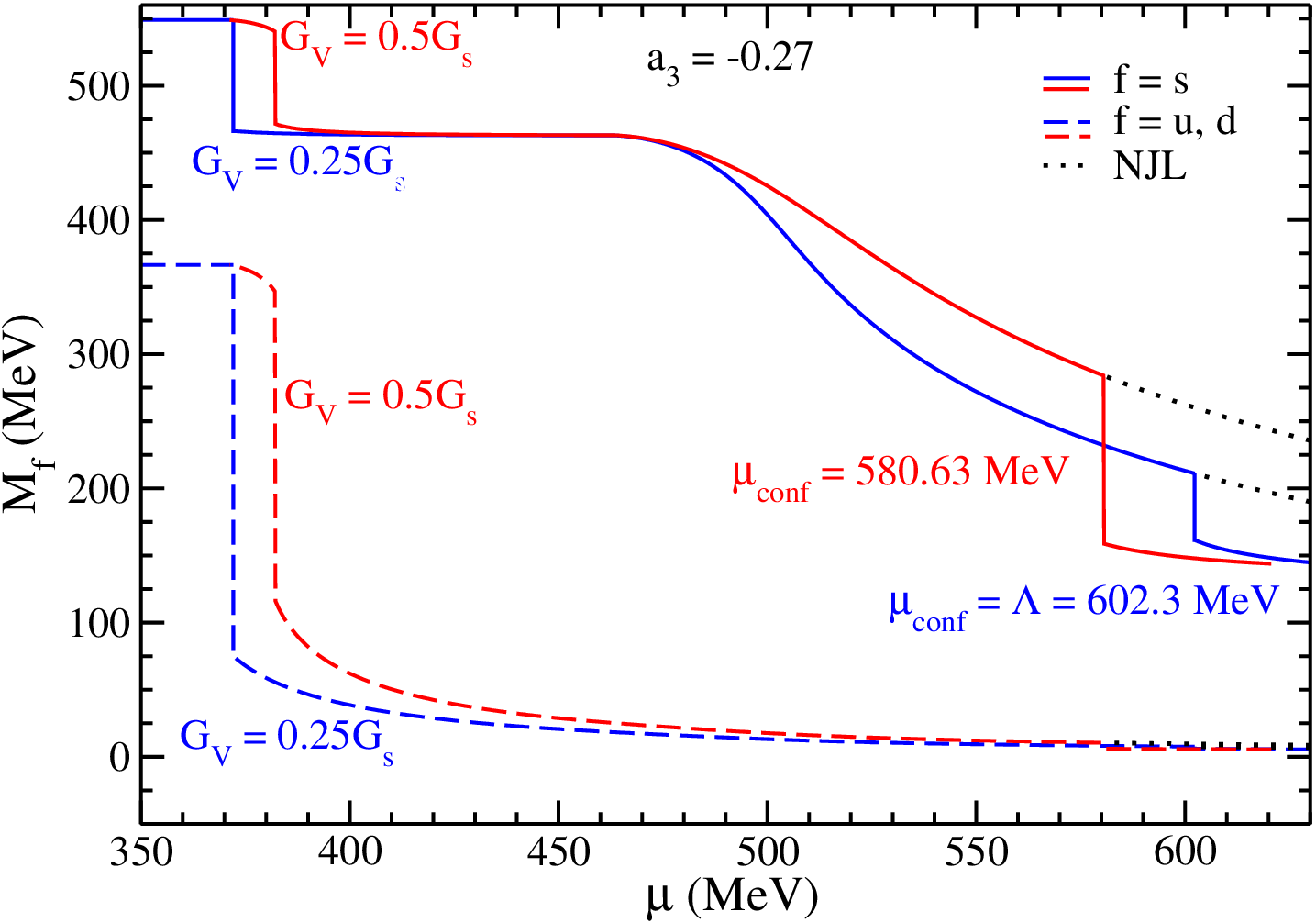}
\caption{The same as in Fig.~\ref{masses}, but for $a_3=-0.27$.} 
\label{masses2}
\end{figure}

As already pointed out, the effect of the traced Polyakov loop depends on $G_V$ and $a_3$. In Fig.~\ref{masses2} for $a_3=-0.27$ and $G_V=0.25G_s$, we verify that $\mu_{\mbox{\tiny conf}}$ is equal to the cutoff parameter~$\Lambda$ resulting a smaller reduction of $M_s$ with respect to the NJL model, differently of what is seen in Fig.~\ref{masses} in the previous case. For the sake of comparison, we also show the results concerning the parametrization presenting $G_V=0.5G_s$ (value obtained from the Fierz transformation of the current-current interaction). Notice that the decrease of the reduction in $M_s$ is also observed for this case.

\subsection{Charge neutral system of quarks and leptons in weak equilibrium within hybrid stars}

In this section we proceed to show how the confinement effects described by the traced Polyakov loop affect the charge neutral system of quarks and leptons (muons and massless electrons) in weak equilibrium. This  study is relevant in the context of quark stars or  hybrid stars~\cite{leupold,stars1,stars2,stars3,stars4,stars5,stars6,stars7}, constructed from a particular quark model at zero temperature and at high density regime. This is a system fulfilling the following conditions for the quarks and leptons chemical potentials: $\mu_s=\mu_d=\mu_u+\mu_e$, and $\mu_e=\mu_\mu$. It is possible to write $\mu_u$, $\mu_d$ and $\mu_s$ in terms of a common quark chemical potential ($\mu$), and in terms of the electron one ($\mu_e$) as
\begin{eqnarray}
\mu_u = \mu - \frac{2}{3}\mu_e, \quad \mbox{and} \quad \mu_d = \mu_s = \mu + \frac{1}{3}\mu_e,
\end{eqnarray}
in which the relation $\mu_B=3\mu$ holds. Furthermore, the requirement of charge neutral quark matter leads to the additional relation to be satisfied
\begin{eqnarray}
\frac{2}{3}\rho_u - \frac{1}{3}\rho_d - \frac{1}{3}\rho_s = \rho_e + \rho_\mu,
\end{eqnarray}
in which $\rho_e$ is the electron density. $\mu_e$ and $\rho_e$ are related to each other through $\rho_e=\mu_e^3/(3\pi^2)$. The muon density is $\rho_\mu=[(\mu_\mu^2 - m_\mu^2)^{3/2}]/(3\pi^2)$, and $m_\mu=105.7$~MeV. In this case, the total energy density and total pressure are given, respectively, by
\begin{eqnarray}
\mathcal{E}_{tq} = \mathcal{E}_{\mbox{\tiny PNJL0}} + \frac{\mu_e^4}{4\pi^2}
+ \frac{1}{\pi^2}\int_0^{\sqrt{\mu_\mu^2-m^2_\mu}}\hspace{-0.6cm}dk\,k^2(k^2+m_\mu^2)^{1/2},\qquad
\label{totaled}
\end{eqnarray}
and
\begin{align} 
P_{tq} & = P_{\mbox{\tiny PNJL0}} + \frac{\mu_e^4}{12\pi^2}
+\frac{1}{3\pi^2}\int_0^{\sqrt{\mu_\mu^2-m^2_\mu}}\hspace{-0.5cm}\frac{dk\,k^4}{(k^2+m_\mu^2)^{1/2} }.
\label{totalp}
\end{align}

\begin{figure}[!htb] 
\centering
\includegraphics[scale=0.33]{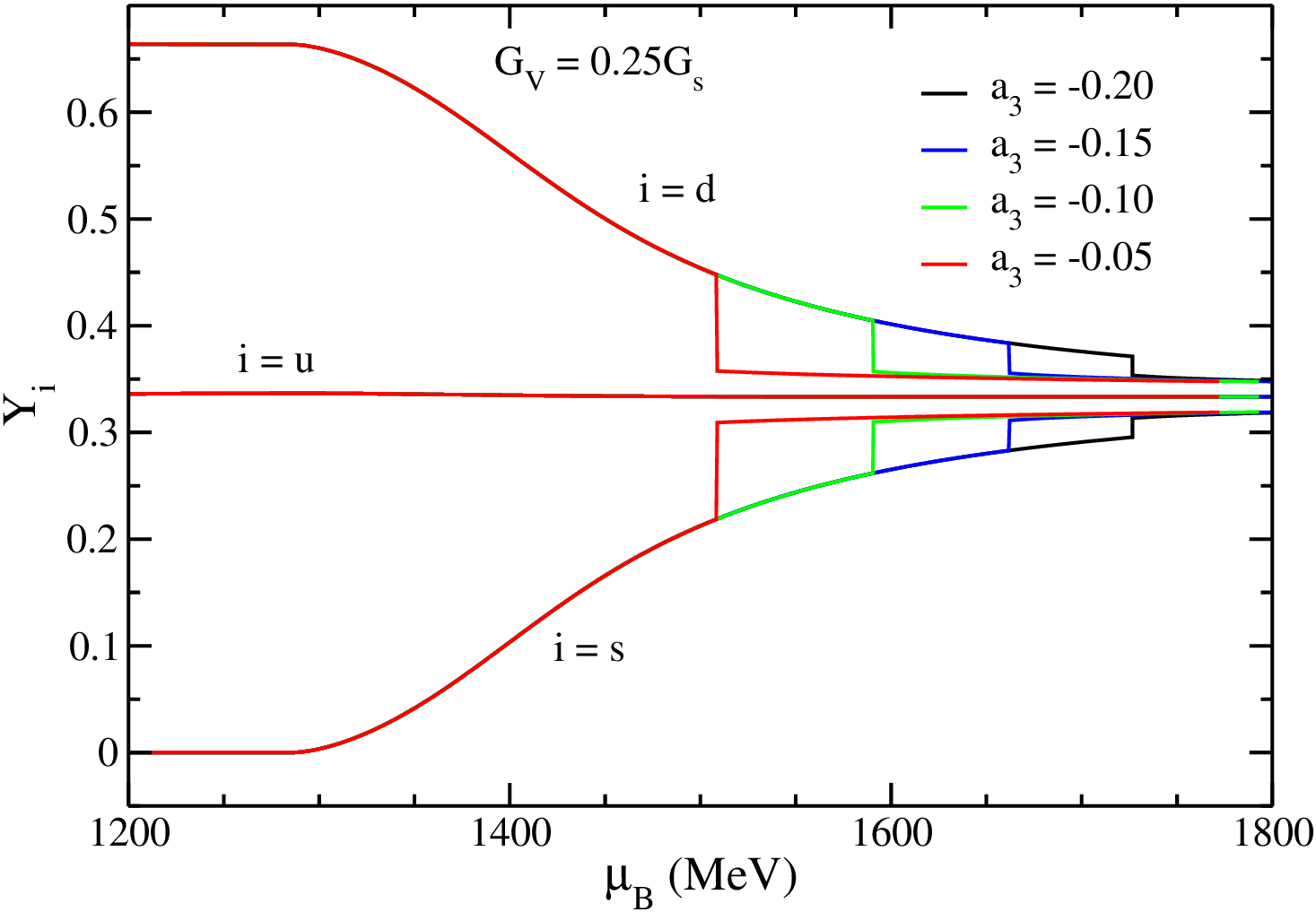}
\caption{Quark fractions as a function of $\mu_B$ for the charge neutral system of quarks and leptons in weak equilibrium. PNJL0 parametrization for $G_V=0.25G_s$ and different values of $a_3$.} 
\label{fractions}
\end{figure}

We start by showing in Fig.~\ref{fractions} the $\mu_B$ dependence of the quark fractions defined as $Y_i=\rho_i/(\rho_u+\rho_d+\rho_s)$, for the SU(3) PNJL0 model with $G_V=0.25G_s$ and different values of $a_3$.
We clearly verify that $d$ quark fraction gradually reduces to a value close to the $u$ quark fraction. Furthermore, at the chemical potential related to the deconfinement transition, point in which $\Phi$ becomes nonvanishing, one observes that this reduction is anticipated by a sharp variation. For the $s$ quark fraction, the opposite is found, namely, $Y_s$ approaches to $Y_u$ with sharp variations produced by the emergence of $\Phi$. In the case of $Y_u$, we find no significant variations along the $\mu_B$ dependence in comparison with both previous cases. From the figure, it is also clear that the critical baryonic chemical potential moves to the direction of increasing $\mu_B$ as $|a_3|$ increases, feature also present in symmetric quark matter, as shown in Table~\ref{taba3}.

As mentioned before, an important application of a quark model available at zero temperature is the description of hybrid stars, in which a hadron-quark phase transition takes place at high density regime. A recent study based on a model-independent analysis, performed in Ref.~\cite{nat}, suggests evidences for massive stars composed by cores of quark matter. Here we take the PNJL0 model to take into account the quark side of the equations of state used to describe stellar matter. For the hadronic side, we use the relativistic mean-field model \mbox{DDH$\delta$} given by the following Lagrangian density
\begin{align}
&\mathcal{L}_{\mbox{\tiny HAD}} = \overline{\psi}(i\gamma^\mu\partial_\mu - M)\psi 
+ \Gamma_\sigma(\rho)\sigma\overline{\psi}\psi 
- \Gamma_\omega(\rho)\overline{\psi}\gamma^\mu\omega_\mu\psi 
\nonumber\\
&-\frac{\Gamma_\rho(\rho)}{2}\overline{\psi}\gamma^\mu\vec{\rho}_\mu\vec{\tau}\psi 
+ \Gamma_\delta(\rho)\overline{\psi}\vec{\delta}\vec{\tau}\psi
+ \frac{1}{2}(\partial^\mu \sigma \partial_\mu \sigma - m^2_\sigma\sigma^2)
\nonumber\\
&- \frac{1}{4}F^{\mu\nu}F_{\mu\nu} 
+ \frac{1}{2}m^2_\omega\omega_\mu\omega^\mu 
-\frac{1}{4}\vec{B}^{\mu\nu}\vec{B}_{\mu\nu}+\frac{1}{2}m^2_\rho
\vec{\rho}_\mu\vec{\rho}^\mu 
\nonumber\\
&+ \frac{1}{2}(\partial^\mu\vec{\delta}\partial_\mu\vec{\delta} 
- m^2_\delta\vec{\delta}^2),
\label{dldd}
\end{align}
where
\begin{eqnarray}
\Gamma_i(\rho) = \Gamma_i(\rho_0)a_i\frac{1+b_i(\rho/\rho_0+d_i)^2}{1+c_i(\rho/\rho_0+d_i)^2},
\label{gamadefault}
\end{eqnarray}
for $i=\sigma,\omega$, and
\begin{eqnarray}
\Gamma_i(\rho)=\Gamma_i(\rho_0)[a_ie^{-b_i(\rho/\rho_0-1)} - c_i(\rho/\rho_0 - d_i)].
\label{gamarho}
\end{eqnarray}
for $i=\rho,\delta$. In Eq.~(\ref{dldd}) $\psi$ is the nucleon field whereas $\sigma$, $\omega^\mu$, $\vec{\rho}_\mu$, and $\vec{\delta}$ are the scalar, vector, isovector-vector, and isovector-scalar fields, that represents mesons $\sigma$, $\omega$, $\rho$, and~$\delta$, respectively. The antisymmetric tensors $F_{\mu\nu}$ and $\vec{B}_{\mu\nu}$ are written as $F_{\mu\nu}=\partial_\nu\omega_\mu-\partial_\mu\omega_\nu$ and $\vec{B}_{\mu\nu}=\partial_\nu\vec{\rho}_\mu-\partial_\mu\vec{\rho}_\nu$. The nucleon rest mass is $M_{\mbox{\tiny nuc}}$, and the mesons masses are $m_\sigma$, $m_\omega$, $m_\rho$, and $m_\delta$. A more simplified version of this model was firstly proposed by J. D. Walecka in Ref.~\cite{walecka} in which the couplings between mesons and nucleons are constants. Here we choose to use a more sophisticated structure for the hadronic model in which such couplings are density dependent, namely, $\Gamma_i(\rho)$, with $\rho$ being the sum of the densities of proton ($\rho_p$) and neutrons ($\rho_n$). In Eqs.~(\ref{gamadefault}) and~(\ref{gamarho}) $\rho_0$ is the nuclear matter saturation density and $\Gamma_i(\rho_0)$, $a_i$, $b_i$, $c_i$, and $d_i$ are free parameters of the model, that along with the nucleon and meson masses, define a specific hadronic parametrization. In particular, we use here the \mbox{DDH$\delta$}~\cite{ddhd} one. It is one of the 263 parametrizations tested against a set of constraints related to symmetric nuclear matter, pure neutron matter, symmetry energy and its slope (evaluated at $\rho=\rho_0$)~\cite{PRC90-055203}. It was also shown that this particular model is capable to generate massive neutron stars with mass around two solar masses~\cite{PRC93-025806}.

Concerning the inclusion of more baryons in the hadronic sector, we remind the reader that some studies point out to a possible suppression of the hyperons population in hadron-quark phase transitions~\cite{stars6,sup} due to the onset of hyperons occurring above the onset of quark matter. Furthermore, the inclusion of hyperons soften the hadronic equation of state with a direct consequence of producing not so massive stars in comparison with the case in which only nucleons are considered. Because of that, and since the main focus of this work is to analyze the thermodynamical structure of a quark model presenting deconfinement effects at zero temperature regime (PNJL0 model), we decide to use a RMF model only with nucleons, avoiding a proliferation of more coupling constants to be fixed in the hadronic sector (hyperons interactions are not completely determined and some assumptions are required).

The Euler-Lagrange equations are used in order to obtain the field equations of the model. In addition, the implementation of the Hartree approximation (no Fock term)~\cite{serot,walecka} in these equations, leading to $\sigma\rightarrow \left<\sigma\right>\equiv\sigma$, $\omega_\mu\rightarrow \left<\omega_\mu\right>\equiv\omega_0$, $\vec{\rho}_\mu\rightarrow \left<\vec{\rho}_\mu\right>\equiv \bar{\rho}_{0(3)}$, and $\vec{\delta}\rightarrow\,\,<\vec{\delta}>\equiv\delta_{(3)}$, allows the determination of the energy-momentum tensor, quantity used to calculate the energy density and pressure of the hadronic model. These last two thermodynamical equations of state are given by
\begin{align}
\mathcal{E}_{\mbox{\tiny HAD}} &= 
\frac{1}{2}m^2_\sigma\sigma^2 
- \frac{1}{2}m^2_\omega\omega_0^2 
- \frac{1}{2}m^2_\rho\bar{\rho}_{0(3)}^2
+ \frac{1}{2}m^2_\delta\delta^2_{(3)}
\nonumber \\
&+\frac{\Gamma_\rho(\rho)}{2}\bar{\rho}_{0(3)}\rho_3 
+\frac{1}{\pi^2}\int_0^{{k_F}_{p}}dk\,k^2(k^2+M_p^{*2})^{1/2}
\nonumber \\
&+\Gamma_\omega(\rho)\omega_0\rho + \frac{1}{\pi^2}\int_0^{{k_F}_{n}}dk\,k^2(k^2+M_n^{*2})^{1/2}
\label{denergdd}
\end{align}
and
\begin{align}
P_{\mbox{\tiny HAD}} &= \rho\Sigma_R(\rho)- \frac{1}{2}m^2_\sigma\sigma^2 +
\frac{1}{2}m^2_\omega\omega_0^2 
+ \frac{1}{2}m^2_\rho\bar{\rho}_{0(3)}^2 
\nonumber \\
&- \frac{1}{2}m^2_\delta\delta^2_{(3)}
+\frac{1}{3\pi^2}\int_0^{{k_F}_{p}}\hspace{-0.5cm}\frac{dk\,k^4}{(k^2+M_p^{*2})^{1/2}}
\nonumber \\
&+\frac{1}{3\pi^2}\int_0^{{k_F}_{n}}\hspace{-0.5cm}\frac{dk\,k^4}{(k^2+M_n^{*2})^{1/2}} 
\label{pressuredd}
\end{align}
with
\begin{align}
\Sigma_R(\rho)&=\frac{\partial\Gamma_\omega}{\partial\rho}\omega_0\rho
+\frac{1}{2}\frac{\partial\Gamma_\rho}{\partial\rho}\bar{\rho}_{0(3)}\rho_3
-\frac{\partial\Gamma_\sigma}{\partial\rho}\sigma\rho_s
\nonumber\\
&-\frac{\partial\Gamma_\delta}{\partial\rho}\delta_{(3)}\rho_{s3},
\end{align}
\begin{align}
\rho_s &=  \rho_{sp} + \rho_{sn}
\nonumber\\
&=\frac{M^*}{\pi^2}\left[\int_0^{{k_F}_{p}}\hspace{-0.5cm}
\frac{dk\,k^2}{\sqrt{k^2+M_p^{*2}}} 
+\int_0^{{k_F}_{n}}\hspace{-0.5cm}
\frac{dk\,k^2}{\sqrt{k^2+M_n^{*2}}}\right],
\end{align}
$\rho_3=\rho_p-\rho_n$, and $\rho_{p,n}={k_F}_{p,n}^3/3\pi^2$. Furthermore, the fields are determined as $\sigma = \Gamma_\sigma(\rho)\rho_s/m^2_\sigma$, $\omega_0 = \Gamma_\omega(\rho)\rho/m_\omega^2$, $\bar{\rho}_{0(3)} = \Gamma_\rho(\rho)\rho_3/2m_\rho^2$, and $\delta_{(3)}=\Gamma_\delta(\rho)\rho_{s3}/m_\delta^2$ with $\rho_{s3}=\rho_{sp}-\rho_{sn}$. The nucleon effective masses read 
\begin{eqnarray}
M^*_{p,n}=M_{\mbox{\tiny nuc}}-\Gamma_\sigma(\rho)\sigma\pm \Gamma_\delta(\rho)\delta_{(3)}
\end{eqnarray}
with~$(-)$ for protons and~$(+)$ for neutrons. Finally, the proton and neutron chemical potentials, obtained through $\mu_{p,n}=\partial\mathcal{E_{\mbox{\tiny HAD}}}/\partial\rho_{p,n}$, are given by
\begin{eqnarray}
\mu_p &=& (k_{Fp}^2+{M_p^*}^2)^{1/2} + \Gamma_\omega\omega_0 +\frac{\Gamma_\rho}{2}\bar{\rho}_{0(3)} +\Sigma_R
\end{eqnarray}
and
\begin{eqnarray}
\mu_n &=& (k_{Fn}^2+{M_n^*}^2)^{1/2} + \Gamma_\omega\omega_0 -\frac{\Gamma_\rho}{2}\bar{\rho}_{0(3)}
+ \Sigma_R,
\end{eqnarray}
respectively. We also implement charge neutrality and chemical equilibrium in the hadronic side by including muons and massless electrons in the system, in the same way it was done in the quark sector. This leads to the following relations: $\rho_p-\rho_e=\rho_\mu$ and $\mu_n-\mu_p=\mu_e$, with $\mu_\mu=\mu_e$. In this case, total energy density and total pressure are constructed as $\mathcal{E}_{tH}=\mathcal{E}_{\mbox{\tiny HAD}}+\mathcal{E}_e+\mathcal{E}_\mu$ and $P_{tH}=P_{\mbox{\tiny HAD}}+P_e+P_\mu$. Moreover, $\mathcal{E}_{e,\mu}$ and $P_{e,\mu}$ can be extracted from Eqs.~(\ref{totaled}) and~(\ref{totalp}). 

\begin{figure}[!htb] 
\centering
\includegraphics[scale=0.32]{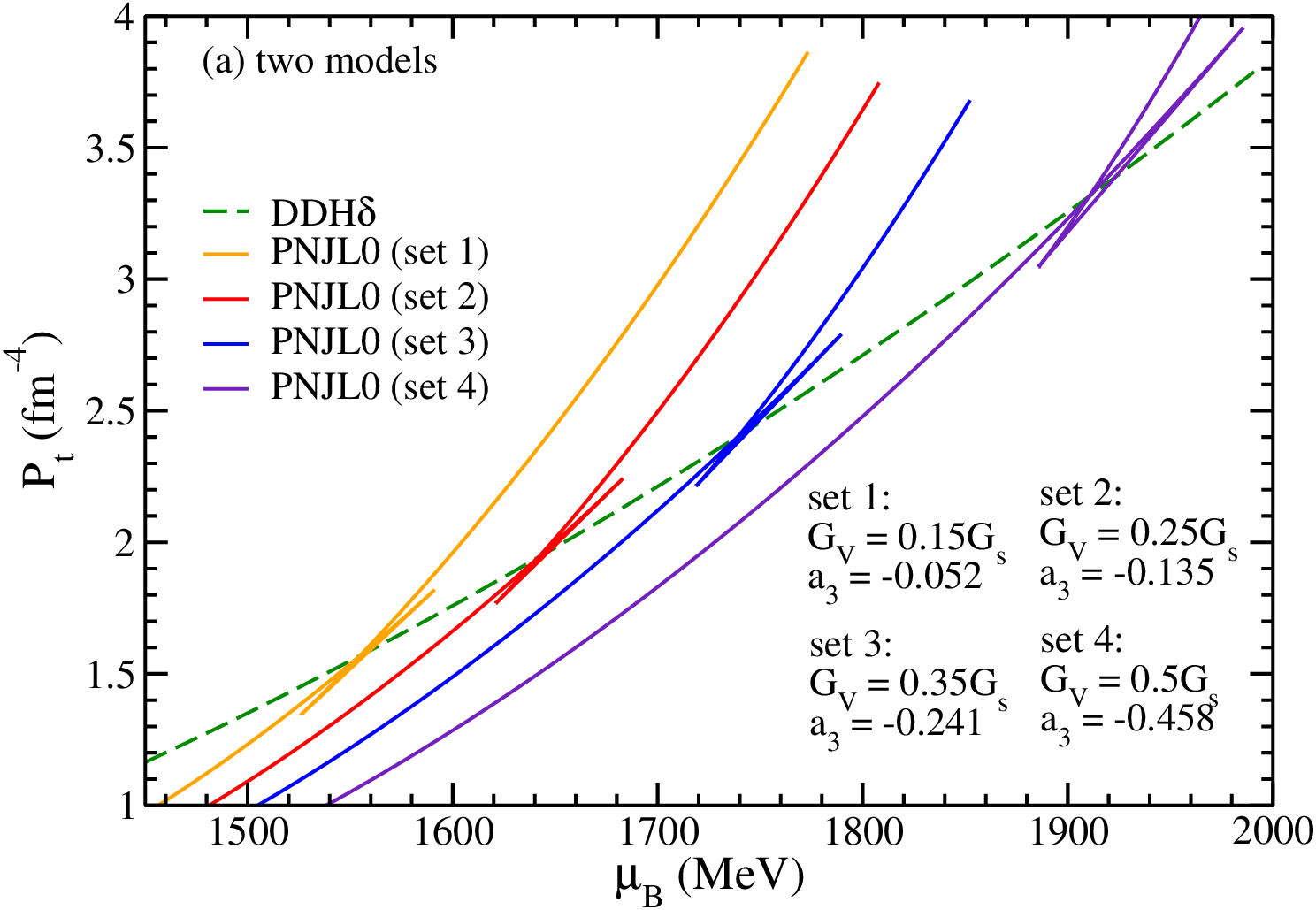}
\includegraphics[scale=0.32]{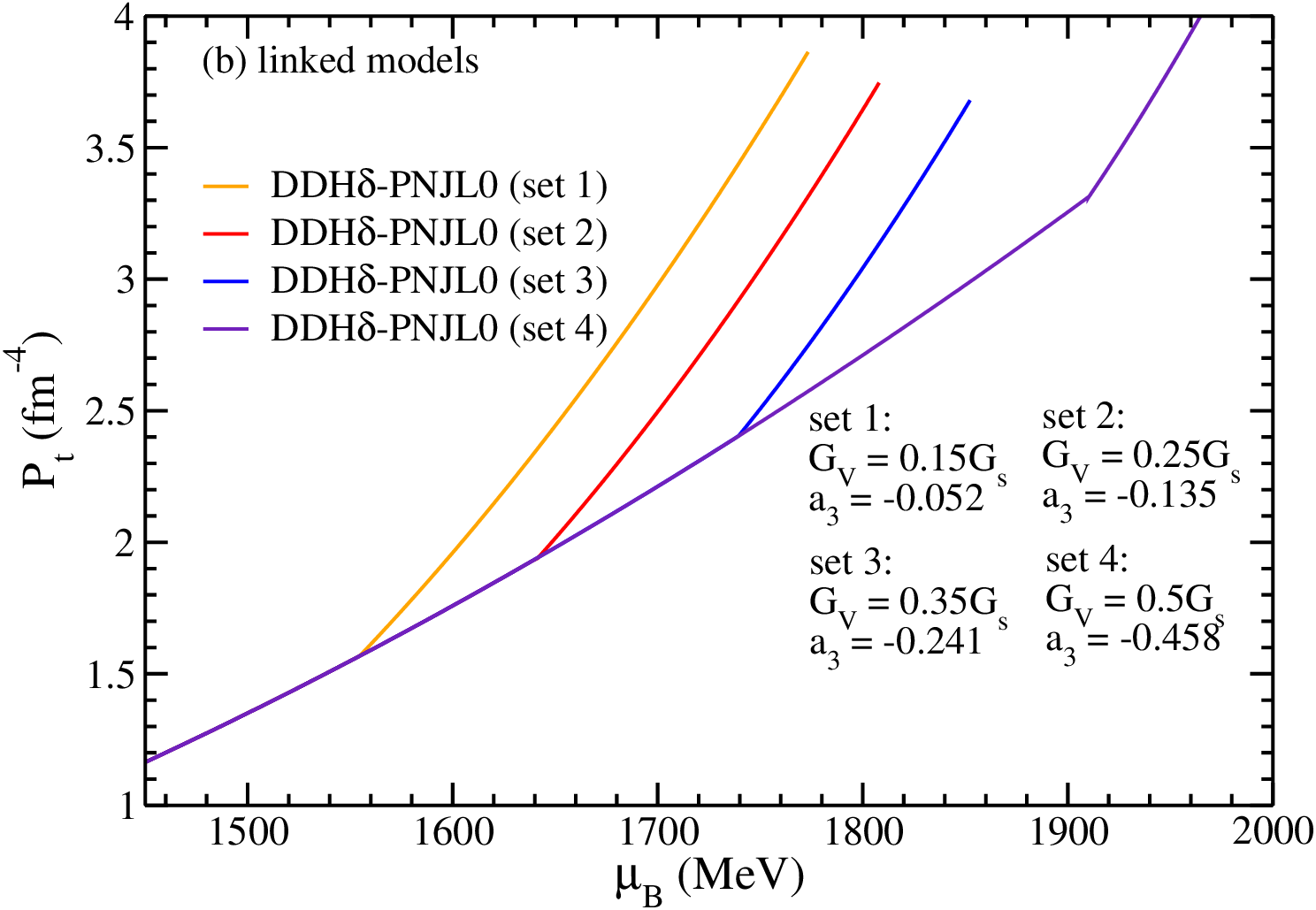}
\caption{Total pressure as a function of $\mu_B$ for (a) two models: \mbox{DDH$\delta$} and different PNJ0 parametrizations, and (b) linked models built through Maxwell construction.} 
\label{pmu}
\end{figure}

In Fig.~\ref{pmu}a we display total pressure as a function of $\mu_B$ for the hadronic model and for three different parametrizations of the PNJL0 quark model. 
In the hadronic side, one has $\mu_B=\mu_n$. For the quark sector we construct the PNJL0 parametrizations as follows: first we choose some values for the ratio $G_V/G_s$, namely, $0.15$, $0.25$, $0.35$ and $0.5$. For each of these values we impose that the point related to the confinement/deconfinement phase transition takes place exactly at the match with the hadronic equation of state. The value of $a_3$ is then determined for this purpose. This procedure, also adopted in Ref.~\cite{pnjl0outro}, generates the linked models presented in Fig.~\ref{pmu}b, in which the entire curves start with the hadronic model until the crossing point, and change to the deconfined quark model thereafter. This is the basis of the Maxwell construction, that imposes equal pressures and chemical potentials at both phases, for a fixed temperature, with a sharp first-order phase transition. The PNJL0 parametrizations constructed from this method present (i) set 1: $G_V/G_s=0.15$, $a_3=-0.052$; (ii) set 2: $G_V/G_s=0.25$, $a_3=-0.135$; (iii) set 3: $G_V/G_s=0.35$, $a_3=-0.241$; and (iv) set 4: $G_V/G_s=0.5$, $a_3=-0.458$. Notice that in set 4, where $G_V/G_s=0.5$, we found a value very close ($a_3=-0.458$) to one used in Ref.~\cite{Schramm} ($a_3=-0.4$).

\begin{figure}[!htb] 
\centering
\includegraphics[scale=0.31]{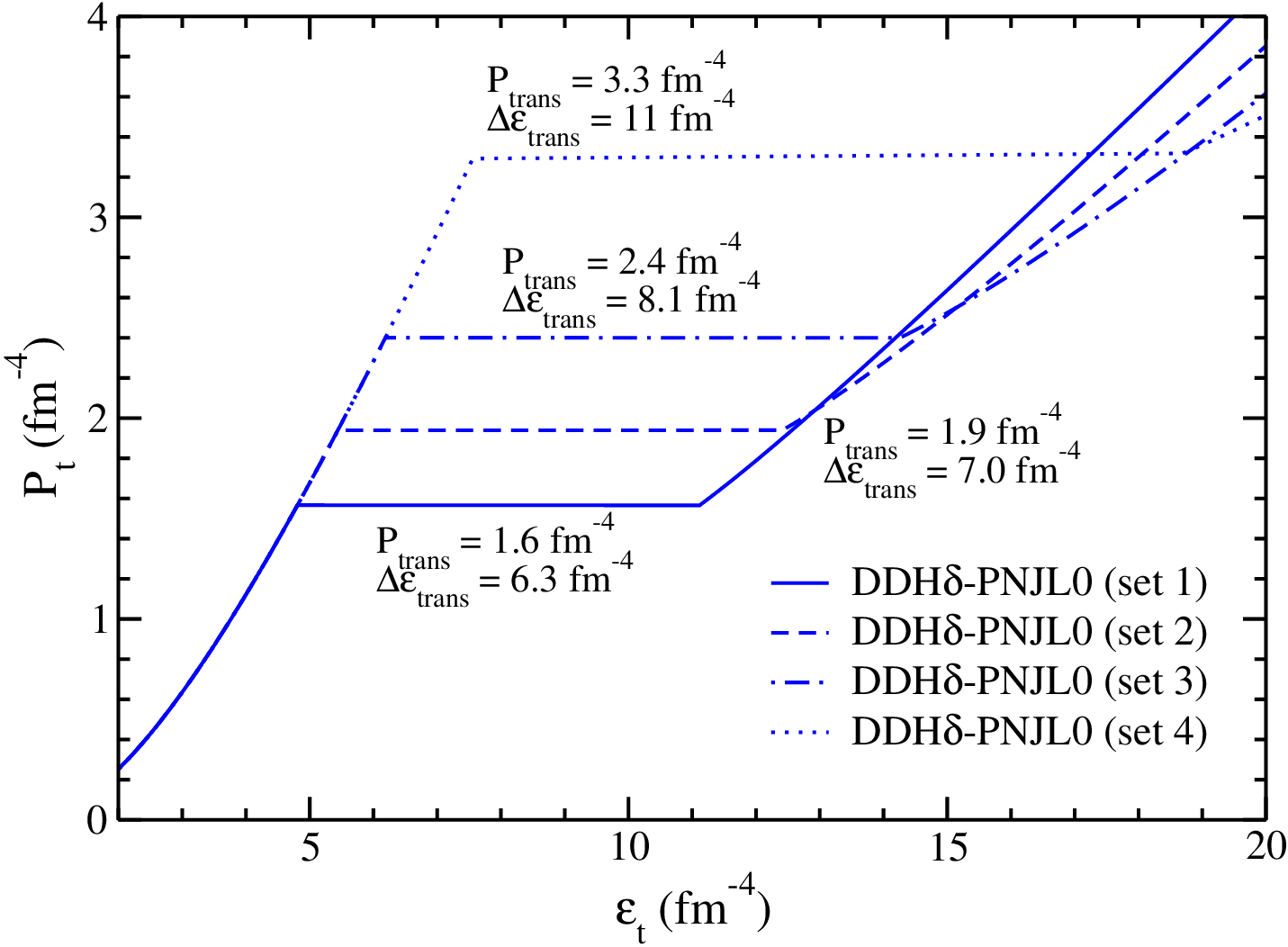}
\caption{Total pressure as a function of total energy density for the linked models. } 
\label{pe}
\end{figure}

In Fig.~\ref{pe} it is depicted how $P_t$ depends on $\mathcal{E}_t$ for the linked models. It is clear that the effect of the first-order phase transition is to establish a plateau in the pressure with a gap in the energy density. The values of the transition pressure ($P_{\mbox{\tiny trans}}$) and the energy density gap at the transition point ($\Delta\mathcal{E}_{\mbox{\tiny trans}}$) for each PNJL0 parametrization are also shown. 
Another interesting feature present in this analysis is that the two quantities, $P_{\mbox{\tiny trans}}$ and $\Delta\mathcal{E}_{\mbox{\tiny trans}}$, increase with $G_V$, or equivalently, with $|a_3|$, both free parameters related to the PNJL0 quark model. The increasing of $P_{\mbox{\tiny trans}}$ means that the hadronic side of the linked model persists at higher values of $\mu_B$. This leads to a decreasing of the quark side, consequently, whilst the increasing of $\Delta\mathcal{E}_{\mbox{\tiny trans}}$ indicates an increasing of the coexistence region of hadrons and (deconfined) quarks.

\begin{figure}[!htb] 
\centering
\includegraphics[scale=0.33]{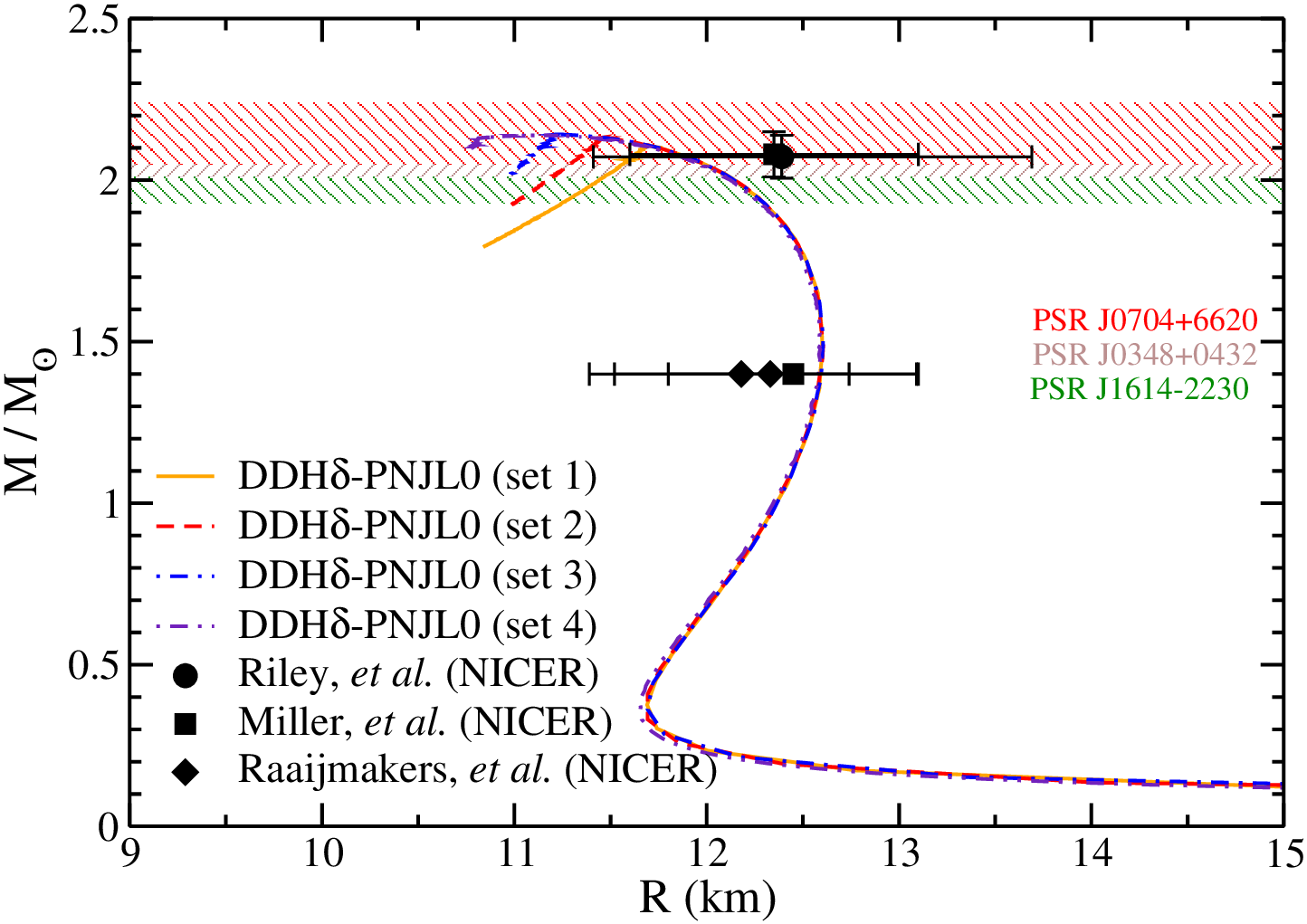}
\caption{Hybrid star mass over $M_\odot$ (solar mass) as a function of its radius for the \mbox{DDH$\delta$-PNJL0} models. Bands extracted from Refs.~\cite{Demorest,Antoniadis,Cromartie}. Circle, squares and diamonds with error bars are related to the NICER data~\cite{a79,a80,a81}.} 
\label{mr}
\end{figure}

In order to describe a spherically symmetric hybrid star of mass~$M$, we solve the Tolman-Oppenheimer-Volkoff (TOV) equations~\cite{tov39,tov39a,glen} given by ($G=c=1$) 
\begin{align}
\frac{dp(r)}{dr}&=-\frac{[\epsilon(r) + p(r)][m(r) + 4\pi r^3p(r)]}{r^2[1-2m(r)/r]},
\label{tov1}
\\
\frac{dm(r)}{dr}&=4\pi r^2\epsilon(r),
\label{tov2}
\end{align}
whose solution is constrained to $p(0)=p_c$ (central pressure) and $m(0) = 0$. At the star surface, one has $p(R) = 0$ and $m(R)\equiv M$, with $R$ defining its radius. The equations of state used as input to solve Eqs.(\ref{tov1})-(\ref{tov2}) are given by total pressure and total energy density of the linked models presented before. Furthermore, at the hadronic side we describe the crust of the star by the model developed by Baym, Pethick and Sutherland (BPS)~\cite{bps} in a density region from $\rho=0.16\times10^{-10}\,\mbox{fm}^{-3}$ to $\rho=0.89\times10^{-2}\,\mbox{fm}^{-3}$. The mass-radius profiles of the hybrid stars obtained from the models studied here are shown in Fig.~\ref{mr}.
In this figure we compare our results with some important observational data. Two of them are related to the mass values of the objects PSR J1614-2230 and PSR J0348+0432 with $M=(1.97\pm 0.04)M_{\odot}$~\cite{Demorest} and $M=(2.01\pm 0.04)M_{\odot}$~\cite{Antoniadis}, respectively, and the recent one related to the MSP J0740+6620 pulsar at $68.3\%$ credible level, namely, $M=2.14{{+0.10}\atop{-0.09}} M_{\odot}$~\cite{Cromartie}. As one can see, the hybrid model is able to produce stars with masses inside these boundaries. For the sake of comparison, we also depict in Fig.~\ref{mr} the recent data extracted from the Neutron Star Interior Composition Explorer (NICER) mission concerning the radius related to the stars of masses equal to $2.08M_\odot$ and $1.4M_\odot$. Such points are given by $M=1.4M_\odot$ with $R=(12.45\pm0.65)$~km~\cite{a79}, $M=(2.08\pm0.07)M_\odot$ with $R=(12.35\pm0.75)$~km~\cite{a79} (squares); $M=2.072_{-0.066}^{+0.067}M_\odot$ with $R=12.39_{-0.98}^{+1.30}$~km~\cite{a80} (circle); $M=1.4M_\odot$ with $R=12.33_{-0.81}^{+0.76}$~km~\cite{a81}, $M=1.4M_\odot$ with $R=12.18_{-0.79}^{+0.56}$~km~\cite{a81} (diamonds). We also observe that hybrid star mass-radius curve is essentially dominated by the hadronic phase, leaving little room to the detailed dynamics from the PNJL0 model, unless for stars with the smallest radius, where the mass is somewhat decreasing by shrinking its size. That effect comes from the softening of the repulsion in the PNJL0 model when the deconfinement phase transition is approached, without the star loosing the stability, as we discuss below.

\begin{figure}[!htb] 
\centering
\includegraphics[scale=0.33]{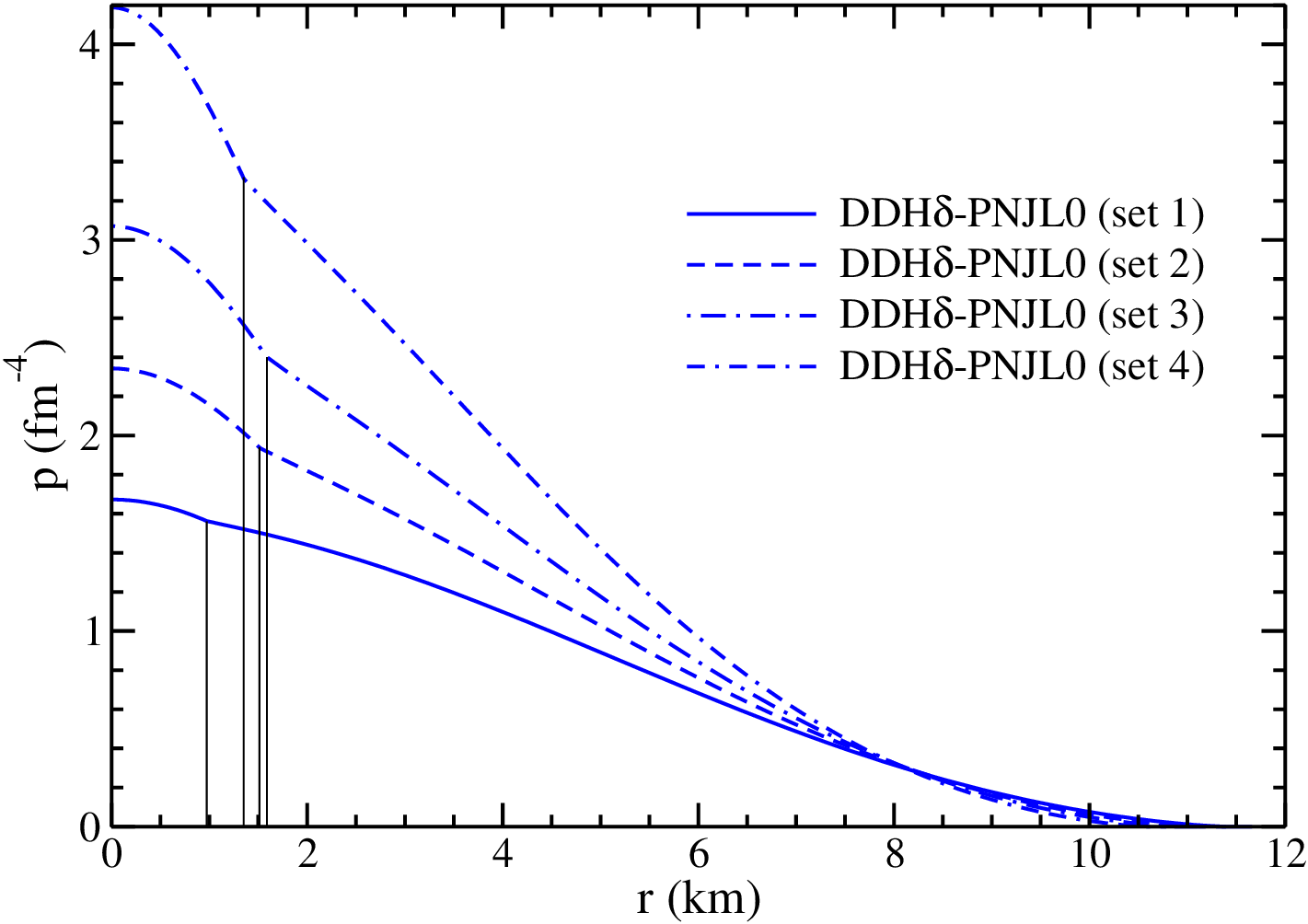}
\caption{Pressure as a function of the radial coordinate for the $M=2.098M_\odot$ star obtained from the \mbox{DDH$\delta$-PNJL0} model (sets 1, 2, 3 and 4). The vertical lines mark the kinks related to the quark cores of each parametrization.}
\label{pr}
\end{figure}

As an illustration, we plot in Fig.~\ref{pr} how the pressure depends on the radial coordinate for a particular hybrid star constructed from the four sets of the \mbox{DDH$\delta$-PNJL0} model. In this case we chose the $M=2.098M_\odot$ star.
Notice that this particular star is predicted by all the sets considered here. In this figure, the kinks identify the quark cores of the star with radius $R_q\sim 1.0$~km for set~1. For sets~2 and~3 we found practically the same result of $R_q\sim 1.5$~km and $R_q\sim 1.6$~km, respectively. Finally, set~4 predicts a quark core of $R_q\sim 1.3$~km. In our approach, the core is composed by deconfined quarks. On the other hand, hadrons are the degrees of freedom of the star from $r=R_q$ to $r=R$.

As a last remark of our results concerning the mass-radius diagrams, we observe that for each  parametrization constructed, namely, sets 1 to 3 of the \mbox{DDH$\delta$-PNJL0} hadron-quark model, one identifies linear branches in which $M$ decreases as $R$ decreases. From the investigation of the static condition for the star's stability, this kind of branch would indicate unstable configurations~\cite{harrison,meltzer}. However, a dynamical analysis of the star stability can give further insights   by verifying its response to small radial perturbations, see for instance Refs.~\cite{rad1,rad2,rad3,rad4,rad5,rad6,rad7} for details. By using this criterion, stars are said stable if such perturbations produce well defined oscillations. On the other hand, an indefinite increasing of the perturbations amplitude characterizes unstable stars. The set of differential equation that need to be solved, coupled to the TOV ones, are given by~\cite{rad1,rad2,rad3,rad4,rad5,rad6,rad7}
\begin{align}
\frac{d\xi}{dr} = -\frac{1}{r}\left(3\xi + \frac{\Delta p}{\Gamma p}\right) - \frac{dp}{dr}\frac{\xi}{(p+\epsilon)}
\end{align}
and
\begin{align}
\frac{d\Delta p}{dr} &= \xi\left[\omega^2e^{\lambda-\nu}(p+\epsilon)r - 4\frac{dp}{dr}\right]
\nonumber\\
&+\xi\left[\left(\frac{dp}{dr}\right)^2\frac{r}{p+\epsilon}-8\pi e^\lambda(p+\epsilon)pr\right]
\nonumber\\
&+\Delta p\left[\frac{dp}{dr}\frac{1}{p+\epsilon}-4\pi(p+\epsilon)re^\lambda\right],
\end{align}
with $e^\lambda=1-2m(r)/r$, $d\nu/dr=-2(dp/dr)(p+\epsilon)^{-1}$, and $\Gamma=(\rho/p)(dp/d\rho)$. The relative radial displacement and the perturbation of the pressure, $\xi=\Delta r/r$ and $\Delta p$, respectively, are assumed to have a time dependence of $e^{i\omega t}$ with $\omega$ being the eigenfrequency. 

 Furthermore, other important aspect that must be mentioned is the kind of phase transition chosen. In stars with two phases, transitions are classified as ``{\it fast}'' or ``{\it slow}'' depending on the timescale of the transition compared to the timescale of the fundamental oscillation~\cite{rad0}. As we can see in Refs.~\cite{rad3,rad4,rad2}, slow transitions have important consequences in the microscopic structure of compact stars with two different phases. It is well known that in stable stars the fundamental frequency has to satisfy the condition of $\omega^2 > 0$, and the last stable star occurs at $\omega = 0$. In general, this property coincides with the maximum mass in the mass-radius diagram. However, when a star presents two phases and the transition is slow, the last stable star ($\omega = 0$) can occur after the point of maximum mass, as we can verify in the results shown in Ref~\cite{rad2}, for instance. By following this approach, we verify that all branches analyzed in the parametrizations shown in Fig.~\ref{mr} are stable, i.e., we found no points in which $\omega=0$.


\section{Summary and concluding remarks}
\label{secsummary}

In this work we generalize the study performed in Ref.~\cite{epjc}, where we have proposed a version of the Polyakov-Nambu-Jona-Lasino that exhibits effects of the confinement/deconfinement phase transition at zero temperature regime for the two-flavor case. We improve our analysis in order to take into account the strangeness in the dense system by introducing the dynamics of the strange quark $s$, whose current mass is considered as $m_s=140.7$~MeV. We start with the Nambu-Jona-Lasinio  Lagrangian density model with the 't Hooft interaction included. We recall the NJL grand canonical potential written in terms of the strengths of the scalar, vector and quark mixture channels, namely, $G_s$, $G_V$ and $K$, respectively. The implementation of the infrared quark-gluon dynamics is performed by making these couplings dependent on the traced Polyakov loop as follows, $G_s \rightarrow \mathcal{G}_s(G_s,\Phi) = G_s( 1-\Phi^2 )$, $G_V \rightarrow \mathcal{G}_V(G_V,\Phi) = G_V( 1-\Phi^2 )$, and $K \rightarrow  \mathcal{K}(K,\Phi) = K( 1-\Phi^2)$. The motivation for this procedure is to ensure vanishing quark interactions at~$\Phi=1$ (deconfined phase). By using these new functions $\mathcal{G}_s(K,\Phi)$, $\mathcal{G}_V(K,\Phi)$ and $\mathcal{K}(K,\Phi)$, it is possible to define a new Polyakov potential now with strangeness included in the back reaction (quarks and gluons affecting each other). 

We also analyze how the confinement/deconfinement phase transition takes place from the study of the minima of $\Phi$ at a fixed (common) quark chemical potential. In this investigation, we focus on the case of symmetric quark matter, namely, the system presenting $\mu_u=\mu_d=\mu_s$. The same study applied to other values of~$\mu$ generates a typical first order phase transition for the order parameter $\Phi$ as a function of $\mu$. Therefore, the system presents two phase transitions, namely, the one defined by $\Phi$, and the other related to the broken/restored chiral symmetry phase transition, in which the quark condensates are the order parameters. This last one is already presented in the original NJL model. The region between these two transitions can be identified, as in the SU(2) case, as the quarkyonic phase, defined here as the phase in which the light quarks present restored chiral symmetry but are still confined ($\Phi=0$). 

Another interesting feature of the SU(3) version of the PNJL0 model is the strong reduction of the constituent strange quark mass ($M_s$) at the  deconfinement phase transition. We verify that $M_s$ undergoes a reduction of almost $50\%$ for the parametrization in which $G_V=0.25G_s$ and $a_3=-0.1$. The dynamics of the traced Polyakov loop at $T=0$ favors the restoration of chiral symmetry even for the strange quark. This effect is not present in the original SU(3) NJL model since $\Phi$ is always vanishing in this case.

We also implement the chemical equilibrium and charge neutrality in the SU(3) PNJL0 model in order to apply it in the description of hybrid stars. For the hadronic side we use a density dependent model and match both equations of state exactly at the point where quarks are deconfined. Different PNJL0 parametrizations were constructed by changing the values of $G_V$ and $a_3$. We  verify that the increasing of both, $G_V$ and $|a_3|$, increases the transition pressure and the gap in the energy density of the hadron-quark phase transition. Finally, we generate the mass-radius profiles determined by the parametrizations of the PNJL0 model (different $G_V$ and~$|a_3|$ values) coupled to the hadronic model. Our results indicate that it is possible to construct hybrid stars compatible with some observational data. In particular, the recent ones proposed by the NICER mission~\cite{a79,a80,a81}.

\section*{ACKNOWLEDGMENTS}
This work is a part of the project INCT-FNA proc. No. 464898/2014-5. It is also supported by Conselho Nacional de Desenvolvimento Científico e Tecnológico (CNPq) under Grants No. 406958/2018-1, 312410/2020-4 (O.L.), No. 433369/2018-3 (M.D.), and 308486/2015-3 (T.F.). We also acknowledge Funda\c{c}\~ao de Amparo \`a Pesquisa do Estado de S\~ao Paulo (FAPESP) under Thematic Project 2017/05660-0 (O.L., M.D., C.H.L, T.F) and Grant No. 2020/05238-9 (O.L., M.D., C.H.L). O.A.M. thanks for the fellowship provided by CNPq - Brazil.

\end{document}